%% file: main.tex
\pdfoutput=1

\documentclass[journal]{IEEEtran}
\ifCLASSINFOpdf
\else
\fi
\usepackage{tabularx}
\usepackage{diagbox}
\usepackage{graphicx}
\usepackage{xcolor}
\usepackage{soul}
\usepackage{hyperref}
\usepackage{float}
\usepackage{cleveref}
\crefformat{section}{\S#2#1#3} % see manual of cleveref, section 8.2.1
\crefformat{subsection}{\S#2#1#3}
\crefformat{subsubsection}{\S#2#1#3}
\input{macros}

\iflong
\usepackage[backend=biber,style=ieee,sorting=ynt,doi=true,isbn=true,date=year]{biblatex}
\else
\usepackage[backend=biber,style=ieee,sorting=ynt,url=false,doi=false,isbn=false,date=year,maxnames=4, abbreviate=true]{biblatex}

\fi
\DeclareUnicodeCharacter{0301}{*************************************}

\begin{document}
% Linebreaks can be used within to get better formatting as desired.
\title{Privacy-Protecting Techniques for Behavioral Biometric Data: A Survey} %\thanks{Identify applicable funding agency here. If none, delete this.}

\author{
\IEEEauthorblockN{Simon Hanisch\IEEEauthorrefmark{1}, Patricia Arias-Cabarcos\IEEEauthorrefmark{3}\IEEEauthorrefmark{2}, Javier Parra-Arnau\IEEEauthorrefmark{2}, Thorsten Strufe\IEEEauthorrefmark{1}\IEEEauthorrefmark{2}}\\
\IEEEauthorblockA{
\textit{Centre for Tactile Internet with Human-in-the-Loop (CeTI), TU Dresden\IEEEauthorrefmark{1}}
%\textit{Technical University Dresden}\\
%Dresden, Germany\\
%first\_name.last\_name@tu-dresden.de
}\\
\and
%\IEEEauthorblockN{2\textsuperscript{nd} Patricia Arias-Cabarcos}\\
\IEEEauthorblockA{
\textit{KASTEL Security Research Labs, KIT\IEEEauthorrefmark{2}}
}\\
\and
\IEEEauthorblockA{
\textit{Human centered IT-Security, Paderborn University\IEEEauthorrefmark{3}}}\\
\and
\IEEEauthorblockA{
%\textit{Karlsruhe Institute of Technology}\\
%Karlsruhe, Germany\\
first\_name.last\_name@tu-dresden.de || pac@mail.upb.de || first\_name.last\_name@kit.edu
}}
\markboth{Journal of \LaTeX Class Files,~Vol.~14, No.~8, August~2015}%
{Shell \MakeLowercase{\textit{et al.}}: Bare Demo of IEEEtran.cls for IEEE Journals}



\maketitle

\begin{abstract}
Our behavior —the way we talk, walk, act or think— is unique and can be used as a biometric trait. It also correlates with sensitive attributes like emotions and health conditions. Hence, techniques to protect individuals’ privacy against unwanted inferences are required, if such data is planned to be processed. To consolidate knowledge in this area, we systematically review applicable anonymization techniques. We taxonomize and compare existing solutions regarding privacy goals, conceptual operation, advantages, and limitations. We review anonymization techniques for the behavioral biometric traits of voice, gait, hand motions, eye-gaze, heartbeat (ECG), and brain activity (EEG). Our analysis shows that some behavioral traits (e.g., voice) have received much attention, while others (e.g., eye-gaze, brain activity) are mostly neglected. We also find that the evaluation methodology of behavioral anonymization techniques can be further improved.
\end{abstract}


% \fi
\IEEEpeerreviewmaketitle

%%%%%%%%%%%%%%%%%%%%%%%%%%%%%%%%%%%%%%%%%%%%%%%%%%%%%%%%%%%%%
%Introduction
%%%%%%%%%%%%%%%%%%%%%%%%%%%%%%%%%%%%%%%%%%%%%%%%%%%%%%%%%%%%%
\section{Introduction}
\input{sections/intro}

%%%%%%%%%%%%%%%%%%%%%%%%%%%%%%%%%%%%%%%%%%%%%%%%%%%%%%%%%%%%%
%BACKGROUND
%%%%%%%%%%%%%%%%%%%%%%%%%%%%%%%%%%%%%%%%%%%%%%%%%%%%%%%%%%%%%
\section{Background}\label{sec:background}
\input{sections/background}

%%%%%%%%%%%%%%%%%%%%%%%%%%%%%%%%%%%%%%%%%%%%%%%%%%%%%%%%%%%%%
%Related Work
%%%%%%%%%%%%%%%%%%%%%%%%%%%%%%%%%%%%%%%%%%%%%%%%%%%%%%%%%%%%%
\subsection{Related Surveys} 
\input{sections/related_work}

%%%%%%%%%%%%%%%%%%%%%%%%%%%%%%%%%%%%%%%%%%%%%%%%%%%%%%%%%%%%%
%Methodology
%%%%%%%%%%%%%%%%%%%%%%%%%%%%%%%%%%%%%%%%%%%%%%%%%%%%%%%%%%%%%
\subsection{Methodology}
\input{sections/methodology}

%%%%%%%%%%%%%%%%%%%%%%%%%%%%%%%%%%%%%%%%%%%%%%%%%%%%%%%%%%%%%
%Behavioral Data Applications and Privacy Concerns
%%%%%%%%%%%%%%%%%%%%%%%%%%%%%%%%%%%%%%%%%%%%%%%%%%%%%%%%%%%%%
\section{Behavioral Data Applications and Privacy Concerns}\label{sec:apps_and_privacy}
\input{sections/apps_and_privacy}

%%%%%%%%%%%%%%%%%%%%%%%%%%%%%%%
% TAXONONY 
%%%%%%%%%%%%%%%%%%%%%%%%%%%
\section{A Taxonomy of Solutions for Behavioral Data Privacy}\label{sec:taxonomy}
\input{sections/taxonomy}

\section{Anonymization Techniques}\label{sec:techniques}
We organize the surveyed techniques according to the behavioral biometric trait they seek to protect. We start with voice as it is the most important trait, then we move on to gait, hand motions, eye-gaze, heartbeat, and brain activity. For each of the traits, we analyze their utility, threat space, anonymization techniques, and evaluation methodology.

\subsection{Voice}\label{sec:voice}
\input{sections/voice}

\subsection{Gait}\label{sec:gait}
\input{sections/gait}

\subsection{Hand Motions}\label{sec:hand_motions}
\input{sections/hand_Motions}

\subsection{Eye-Gaze}\label{sec:eye}
\input{sections/eye_Gaze}

\subsection{Heartbeat}\label{sec:heart}
\input{sections/heartbeat}

\subsection{Brain Activity}\label{sec:eeg}

\input{sections/brain_Activity}

%%%%%%%%%%%%%%%%%%%%%%%%%%%%%%%%%%%%%%%%%%%%%%%%%%%%%%%%%%%%%%%%

\section{Discussion}\label{sec:discussion}
\input{sections/overview_table}
\input{sections/discussion}

\section{Concluding Remarks}
\label{sec:conclusion}

Anonymizing behavioral biometric data is an important task for protecting people's privacy. In our literature review, we found many different behavioral traits that need to be considered and developed a taxonomy to classify the anonymization techniques that can be applied to them by the type of data transformation they perform. While voice anonymization is already a research field with many available solutions, most behavioral biometric traits only got little attention in the literature and therefore anonymizing them remains an open research question. We further found that most anonymization techniques are only evaluated rudimentarily with the assumption of a weak attacker. Improving the evaluation methodology is therefore another open research question. Lastly, we find that the temporal aspect of the data was mostly neglected, both for offering privacy for data streams and for anonymizing the data.

\section*{Acknowledgment}

This work has been supported by the German Research Foundation (DFG, Deutsche Forschungsgemeinschaft) as part of Germany's Excellence Strategy -- EXC 2050/1 -- Project ID 390696704 -- Cluster of Excellence ``Centre for Tactile Internet with Human-in-the-Loop" (CeTI) of Technische Universit{\"a}t Dresden, and by funding of the Helmholtz Association through the KASTEL Security Research Labs (HGF Topic 46.23).

\ifCLASSOPTIONcaptionsoff
  \newpage
\fi



%\bibliographystyle{IEEEtran}
%\bibliographystyle{IEEEtranN}
%\bibliography{newbiblio}
\printbibliography

%\end{IEEEbiography}


%\newpage



%\vfill


\end{document}

%% file: macros.tex
\newcommand{\cont}{continuous conversion}
\newcommand{\discrete}{discrete conversion}
\newcommand{\feature}{feature removal}
\newcommand{\coarse}{coarsening}
\newcommand{\noise}{noise injection}
\newcommand{\perturb}{random perturbation}

\newif\iflong
\newif\ifcomments

\longtrue
\commentsfalse

\ifcomments
\newcommand{\tc}[1]{\textcolor{blue}{\textbf{TS:} #1}}
\newcommand{\sh}[1]{\textcolor{orange}{\textbf{SH:} #1}}
\newcommand{\pc}[1]{\textcolor{magenta}{\textbf{PC:} #1}}
\newcommand{\jp}[1]{\textcolor{violet}{\textbf{JP:} #1}}
\newcommand{\del}[1]{\textcolor{red}{#1}}
\newcommand{\new}[1]{\textcolor{teal}{#1}}
\else
\newcommand{\tc}[1]{}
\newcommand{\sh}[1]{}
\newcommand{\pc}[1]{}
\newcommand{\jp}[1]{}
\newcommand{\del}[1]{}
\newcommand{\new}[1]{#1}
\fi

%% file: sections/intro.tex
%% -*-latex-*-

The ongoing digital transformation is leading to an increasingly comprehensive data collection on citizens. 
Ever improving peripherals, like augmented reality (AR)/virtual reality (VR) goggles, motion capturing suits and gloves, force-feedback input devices, sensor-rich cell phones, smart watches, and other wearables drastically increase the coverage and resolution at which biometrics and behavioral data of individuals become available for processing.

A large amount of such data is shared knowingly, when users post their latest achievements, photos, or opinions on products and current affairs. 
A much larger amount is collected unnoticed, when individuals browse Web pages, use location services and similar apps, or simply enter smart spaces that are enriched with anything from voice assistants to CCTV cameras.

The corresponding behavioral data is highly descriptive of the captured individual and it reveals a multitude of attributes. 
They contain strong indicators for routines, habits, and also medical conditions and `tics'. 
Known correlations between physiological features and medical conditions include the detection of depression or consumption of anti-depressants in facial pictures, detection of organ insufficiencies due to the coloration of eyes (hepatitis), or skin (alcohol abuse \cite{bruno13study}, general fitness \cite{perret20skin}, and others). 
A large number of studies have also reported correlations between behavioral data and psychological traits as well as characteristics. 
Behavioral data can also be used to uniquely identify individuals. 
Prominent examples across the spectrum include identifying personal traits and characteristics from social media feeds \cite{kosinski13private}, identifying users by their mobility patterns \cite{montjoye13unique}, and web-browsing behavior \cite{deusser2020browsing}.
Gait very prominently has been used to identify individuals~\cite{yovel_recognizing_2016,wan2018survey}, and it obviously reveals individual attributes like age, gender, and physiological conditions~\cite{pollick_gender_2005,troje_decomposing_2002}.
%This also increases capabilities for credit and social scoring, based on aggregated digital dossiers.

Preserving the privacy, and ultimately the dignity of individuals who come in the range of sensors and are captured in their behavior requires more sophisticated approaches than removing direct identifiers (IP address, social security number (SSN), blurring a face) or intuitive quasi identifiers (gender, age, ethnicity) in databases.
Note, that the behavioral data captured from humans has both temporal dependencies, as it is captured as a time-series, and physiological dependencies, as human bodies must adhere to both their physiological and general physical limitations.
Due to the strong dependency between observations and to the physiological and physical dependencies, the efficacy of randomized, perturbative anonymization also must be critically reviewed, as the dependencies might be used to infer the identifiable information that the anonymizations seek to remove. 
Context information and habits being represented as strong signals in the data further complicate effective anonymization.  % I'm thinking of phd comic's prof. smith always wearing his green vest, as an example -- but I guess that none of the approaches we survey would actually effectively anonymize such information? Should we mention this in the conclusion or discussion? -- t

A growing corpus of studies is addressing this challenge of anonymizing behavioral data. They focus on a variety of different human traits, ranging from the voice, over gait, to less prominent examples like gestures, heartbeat, and others.
A systematic review of all these approaches, which bridges the attempts in extracting the shared conceptual and methodological similarities is missing, to the best of our knowledge. Further, we want to highlights both differences as well as roads less traveled.

For this paper, we hence set out to systematize the corresponding literature. 
We are interested in privacy-enhancing technologies (PETs) for scenarios in which behavioral data are collected by or shared with third parties to perform a specific operation. As we are interested more in privacy than confidentiality we do not consider approaches in which an entity encrypts its own data to hide it from access by unintended audiences. 
We are rather interested in approaches that protect from unintended revelation of information contained in data that is collected and shared for a different, explicit purpose \cite{decristofaro21critical}. 
We deem `confidential computing', processing based on homomorphic cryptography, or similar approaches in which the data owner is the only entity that learns anything from the data, out of scope of our analysis.

For our study, we followed Kitchenham's guidelines \cite{systematic} to systematically discover and survey the current state of the art, comprising of 101 distinct studies, extracted from a corpus of 296 initially discovered publications.

We identify common applications that process behavioral data, to extract sensible measures of utility, as well as common privacy threats with corresponding adversary models. 
We define two taxonomies of anonymization approaches the first defined by how the anonymization transform the data and the second by which anonymization goal it seeks to protect. 
\del{We define a taxonomy of anonymization approaches, informed by the related work from the fields of statistical disclosure control and anonymous communications.} Next, we provide a detailed overview of the different anonymization approaches, sorted by the trait they aim to protect. 
We provide insight into the corresponding applications that define \del{utility metric}\new{the utility}, and into the privacy threats, privacy goals, applied anonymization concepts, and the evaluation the corresponding scientists performed, together with the data they chose for their studies. \del{Our main findings are that some traits (e.g., voice) received a lot of research interest while others are mostly neglected (e.g., brain activity, eye-gaze). One reason we found for this is the lack of available datasets for these neglect traits.}

\new{Our main findings are that the underlying concepts that are used for anonymizing behavioral biometric data are independent of the biometric trait. This allows us to identify biometric traits for which specific anonymization concepts have not yet been tested.} Further, we find that the general evaluation methodology for behavioral biometric anonymization implies a weak adversary and must be improved to give a good assessment of the privacy protection.

The rest of the article is organized as follows: \cref{sec:background} describes the background on privacy terminology, as well as the related work and our survey approach. 
\cref{sec:apps_and_privacy} introduces behavioral data, applications, and related privacy concerns. 
We define our taxonomy of concepts in \cref{sec:taxonomy}, and survey the field, sorting anonymization techniques by the trait the authors addressed and the conceptual approach taken, in \cref{sec:techniques}. 
We discuss our insights and general lessons learned in \cref{sec:discussion} and conclude the article with a summary in \cref{sec:conclusion}.

%% file: sections/background.tex
In this section we first review the relevant terminology utilized throughout this work and the existing surveys on anonymization techniques. We then present the methodology we used to perform the systematic literature review.

\subsection{Terminology}
Our use of the term \textbf{privacy enhancement} or \textbf{protection} shall refer to the obfuscation of information from any adversarial observers, including the information or service provider, regardless of whether this obfuscation consists in data access control, encryption, minimization of the data revealed, or data modification, perturbation, partial or full, in any manner. In the most abstract sense, the behavioral information to be protected may be composed of various elements, including links or relationships among several pieces of information.

Another important type of information to be obfuscated is directly a user's \textbf{identity}, by itself or accompanied with behavioral or profile information. The close relation between personal devices (such as smartphones or wearables) and their users makes distinctive features in said devices potentially unique identifiers.
In this respect, we adhere to the terminological convention of regarding \textbf{anonymity} as a particular case of privacy, when the data to be protected, without being direct identifiers\footnote{Direct identifiers allow to unequivocally identify individuals. For example, it would be the case of SSNs or full names. In an data-anonymization process, direct identifiers are always removed in the very first phase.}, may be linked with external information to reidentify the individual to whom the data refer.

In the field of statistical disclosure control (SDC)~\cite{Willenborg01B}, the aim is to protect a microdata set, while ensuring that those data are still useful for researchers. A microdata set is a database whose records contain information at the level of individual respondents. In this field, the concepts of \textbf{identity and attribute disclosure} refer to the goal of an attacker to ascertain either the identity of an individual in the microdata set or the confidential attribute/s thereof.\del{In this work, our interpretation of the terms anonymity and privacy will be in the sense meant by the aforementioned concepts of identity and attribute disclosure, respectively.}

\del{We shall employ the term \textbf{utility} to refer to a quantification of the degree of functionality maintained with respect to that intended by a personalized or information service, despite the implementation of privacy mechanisms that may hide or perturb part of the data, along with the degree of quality of service maintained, despite processing, storage, communication and scalability overheads incurred by such mechanisms. We stress that utility in this context does not refer to user-interface design.}
\new{We shall employ the term \textbf{utility} to quantify the degree of functionality maintained concerning a service for which the behavioral biometric data is intended. The utility is kept despite the implementation of a privacy mechanism that may hide or perturb part of the data which may degrade the quality of the service. We stress that utility in this context does not refer to user-interface design.}

As pointed out above in the introduction, any PET poses a \textbf{trade-off between privacy and functionality}. The optimization of the privacy-functionality (or privacy-utility) trade-off will refer to the design and tuning of PETs in order to maximize privacy for a desired functionality, or vice versa.

%% file: sections/related_work.tex
\label{sec:rw}
Most of the surveys on behavioral data focus on analyzing the uniqueness and suitability of behavioral traits to identify people, comparing the accuracy of different approaches and their applicability. In this line of research, we find surveys covering a range of existing behavioral biometrics for user authentication~\cite{surveyImplicit, surveyBio, mahfouz2017survey, liang2020behavioral}, and others focusing on the review of specific traits, such as gait recognition \cite{wan2018survey},  keystrokes~\cite{banerjee2012biometric, teh2013survey}, eye gaze \cite{katsini2020role},  or brainwave biometrics \cite{gui2019survey}. However, the treatment of privacy issues is limited to mentioning that there is potential for sensitive inferences or identity leaks but there is no in-depth discussion about privacy countermeasures.   

There is an important stream of research on potential privacy attacks to behavioral data focusing on \textbf{attribute inferences}~\cite{bales2016gender, inoue2020gender, Frank2013, buriro2016age},  or dealing with user de-identification (i.e., trying to identify a person by their behavioral data)~\cite{ye2021deep, el2011systematic, dwork2017exposed, henriksen2016re}. Dantcheva et al. \cite{dantcheva2015else}
provide an extensive overview of which sensitive attributes, so called soft biometrics (gender, age, ethnicity, weight, etc.), can be inferred from primary biometrics extracted from image and video data.   This survey highlights that protecting privacy of inferred attributes is an open research challenge.  

While the current literature on behavioral data underscores the need for privacy defences, work on this area is still emerging and scattered. So far no comprehensive view of the problem, existing solutions, and challenges has been carried out yet. Ribaric et al. \cite{ribaric_-identification_2016} review techniques to protect user's visual and multimedia data from attribute inferences and re-identification. Though they include a section on behavioral data protection, it only covers a limited number of traits (voice, gait, and gesture) and anonymization techniques that apply when these data have been captured as video, audio, or images. No other sensors are considered. Also closely related, Nhat Tran et al. \cite{tran2021biometrics} survey biometric template protection techniques, but they do it generally without entering in details of the anonymization needs of behavioral biometrics.
\new{Meden et al.~\cite{9481149} survey PETs which are applied to faces looking at different aspects like the privacy guarantees they give and what of conceptual approaches are chosen. Shopon et al.~\cite{jcp1030024} look at the wider variety of biometric traits including gait and writing style. Their taxonomy focuses on whether the anonymizations hide both identity and attributes of the person or retain some soft biometric features.}

\new{The current reviews of behavioral biometric anonymization look only at one specific trait or only review a few anonymization techniques. Missing is a survey that examines in depth a comprehensive set of traditional and modern types of behavioral traits for which solutions have been proposed, and considering different types of recording sensors and use-cases. Further, a comparison of evaluation approaches across behavioral biometric traits has also not been performed yet. By comparing a large set of behavioral biometric traits similarities and differences between the anonymization approaches become apparent and open research questions can be identified.}

\del{
In our article, we go beyond the state of the art by systematically reviewing research works on behavioral data anonymization techniques, examining a comprehensive set of traditional and modern types of behavioral traits for which solutions have been proposed, and considering different types of recording sensors and use-cases. We categorize and compare existing techniques, analyze their associated evaluation approaches and results, and present a summary of challenges pointing at research directions that need attention in future work.}

%% file: sections/methodology.tex
\label{sec:methodology}
We performed a systematic literature review following Kitchenham’s guidelines~\cite{systematic} to identify relevant studies on privacy techniques for behavioral data\iflong, as it is depicted in Figure~\ref{fig:literature}\else\fi.

\iflong
\begin{figure}[!h]
    \centering
      \includegraphics[width=0.49\textwidth]{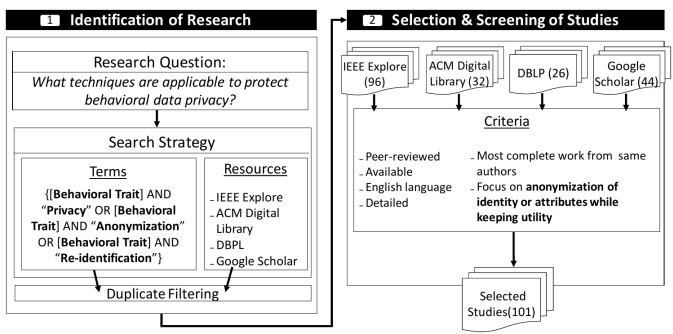}
\caption{
Summary of the procedure for identifying and selecting relevant studies on behavioral data privacy techniques. We first analyzed the literature on biometrics to determine behavioral traits for person identification. We then used these traits as key terms to search for privacy-related publications, following Kitchenham's guidelines for systematic literature reviews~\cite{systematic}. The complete list of behavioral traits we searched includes: brain activity, eye gaze, facial expression, gait, gesture, handwriting, haptic, heartbeat, keystrokes, lip, motion, mouse, thermal, touch, and voice.}
\label{fig:literature} 
\end{figure}
\else\fi    

Our guiding research question is \textbf{``What techniques are applicable to protect behavioral data privacy?''} From this starting point, the goal is to understand how these techniques work, what is the level of protection provided, and what are the limitations and existing open challenges.
To answer these questions, we first explored the literature on biometrics~\cite{surveyImplicit, dantcheva2015else,surveyBio, mahfouz2017survey,yampolskiy2010taxonomy,pfeuffer2019behavioural,ackad2012seamless,griffiths2018privacy} to determine what kind of behavioral traits can be used to identify a person. \new{The complete list of behavioral traits we searched includes: brain activity (also referred to as cognitive biometric), eye gaze, facial expression, gait, gesture, handwriting, haptic, heartbeat, keystrokes, lip, motion, mouse, thermal, touch, and voice.} Next, we used this list of traits combined with the keyword \textbf{``privacy''} and the semantically similar terms \textbf{``anonymization''} and \textbf{``de-identification''}, as search strings in the main academic databases for computer science. Based on these search terms, we compiled works with no constraints on publication date, obtaining a set of 296 papers spanning from 2007 to early 2022, after filtering duplicates. During pre-screening, we  built a taxonomy of privacy solutions and decided to narrow-down the scope of the survey to anonymization techniques focused on protecting the publication of behavioral data from identity and attribute disclosure attacks.  We consider approaches that assume collection, sanitization, and subsequent publishing of data, which must be anonymized but also keep a level of utility to provide behavioral data driven services.
Accordingly, the down-selection of primary studies to be analyzed in this survey considered the following criteria. Documents were excluded if:
\begin{enumerate}
    \item  The publication format was other than peer-reviewed academic journal or conference paper.
    \item The paper could not be retrieved using IEEE Explore, ACM Digital Library, DBLP, or Google Scholar.
    \item The publication language was not English.
    \item Another paper by the same authors superseded the work, in which case the most complete work was considered.
    \item The privacy protection technique was other than identity or attribute anonymization with data utility.
    \item The anonymization approach was described at a high level and not enough details were provided to properly address the guiding research question.
\end{enumerate}

     The search and selection protocol yielded a final corpus of 101 peer-reviewed works on behavioral data anonymization, which we clustered according to the behavioral trait being protected: gait, brain activity, heartbeat, eye gaze, voice,  and hand motions (handwriting, keystrokes, mouse movements, and hand gestures). We found no papers on facial expression, lip, touch, and haptic traits that fulfil our criteria. 
\del{We first describe the different applications of these traits, motivating the need for privacy (Section \ref{sec:apps_and_privacy}). Then, we define a taxonomy for classifying anonymization techniques (Section \ref{sec:taxonomy}). We use this taxonomy to review the papers for each behavioral trait (Section~\ref{sec:techniques}), analyzing  the proposed anonymization technique, its performance, as well as the main advantages and disadvantages.
We then examine and discuss the literature in a consolidated way, identifying overall gaps and future challenges to advance research on behavioral data privacy (Section~\ref{sec:discussion}).}

%% file: sections/apps_and_privacy.tex
Behavioral data can be leveraged to provide valuable services for both users and companies. In this section, we summarize the application model, the main usages of behavioral data and the related emergent privacy issues, which motivate the need for our survey.

\subsection{Behavioral Biometric Data}

Behavioral biometric data are a subclass of biometric data which encompasses all human behavior. While in SDC the columns of a microdata set that should be protected (e.g. name or address) are explicit, for behavioral biometrics as it is not apparent which part of the data is privacy sensitive. As behavioral biometric data are captured from a human, it contains a lot of implicit dependencies between individual data points and across traits. For example the motion of a foot is highly depended on the motion of the corresponding leg. It may immediately imply that a person has been injured, as the behavior exhibits typical patterns of limping, although this attribute has not been made explicit in a field of the record. Another dependency to consider is the temporal dependency between data points as behavioral biometrics are usually captured as a time-series. 
These dependencies make the anonymization of behavioral biometric data challenging as an attacker can use them to reconstruct the clear data and implicit disclosures from the anonymized data.

\subsection{Scenario}

%\iflong
\begin{figure}[!ht]
    \centering
      \includegraphics[width=0.49\textwidth]{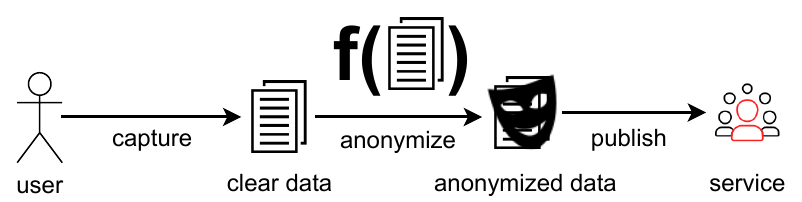}
\caption{The data-publishing scenario of the survey.}
\label{img:publishing}
\end{figure}
%\else\fi

In this survey, we assume a data-publishing scenario (see Figure~\ref{img:publishing}) in which the data are first transformed in a privacy protective manner and then published or shared with a service or application. This also includes involuntary publication, which for example can occur when the biometric templates of an authentication system are leaked. We assume that the utility of the protected, modified data is preserved to the extent that the received service (e.g., a personalized recommendation) is still meaningful and effective.

\subsection{Applications}
\label{sec:applications}
In general, the entire field of \textbf{human computer interaction} captures and processes behavioral biometric data, as each input over time also comprises a behavior. Keystroke patterns and mouse movement are our main input modality for computer systems today, however new input modality such as touch, voice, and gestures are on the rise and will likely become more relevant in the coming years.

Another area where behavioral data are useful is \textbf{healthcare}. Advances in sensors and machine learning techniques enabled the development of applications for activity recognition, fall detection, and remote health monitoring that facilitate caring of elderly, sick, or disabled people and eases diagnosis~\cite{pansiot2007ambient, pogorelc2012automatic, de2017freezing}. Typical collected data are gait and motion information coming from accelerometers and gyroscopes embedded in user devices,  and biosignals like heartbeat or brain activity. This data can be also processed to give health-related feedback to users, for example to guide them through relaxation or to detect and signal cognitive states, such as being stressed, so the user can act on it.

One of the most important and well researched application area of behavioral data is \textbf{biometric recognition} ~\cite{hogben2010enisa,surveyImplicit, surveyBio, mahfouz2017survey}. A person's behavior, such as the way of walking or typing on a keyboard, contain unique inherent patterns that allow for verifying the identity of that person. Given that these patterns can be sensed implicitly while the person interacts with, wears, or carries a device, behavioral biometrics are generally considered more usable than other traditional biometrics like fingerprints~\cite{quest, imperfect}, and therefore a good alternative or complement to password-based authentication. Academic research has shown the feasibility of numerous behavioral traits for user authentication, to name a few: keystroke patterns~\cite{teh2013survey}, gait~\cite{wan2018survey}, touch~\cite{teh2016survey}, mouse movement~\cite{zheng2016efficient}, brain  activity~\cite{gui2019survey}, or even breathing patterns~\cite{chauhan2018breathing, chauhan2017breathprint}. And some of them are already developed in commercial solutions, especially in the financial sector to prevent fraud through detecting behavior anomalies ~\cite{BehavioSec, Typingdna, Nymi, VoiceVault}.

Besides biometric recognition and healthcare, a great deal of behavioral data driven applications are focused on \textbf{personalization}. In this category we find adaptive interfaces and services that change their content or appearance according to the predicted user preferences based on their behavior. Furthermore, personalization can be applied in many areas. To give some examples, behavioral data are used to personalize online games adapting to the player profile for a more satisfactory experience (e.g., adjusting the level of difficulty) \cite{zohaib2018dynamic}, in recommender systems to suggest online content or advertisements \cite{reddy2019content}, or in education to taylor the learning experience to the student mental state (level of attention, stress, etc.) \cite{paul2019eye}.

\subsection{Utility}
Depending on each application the behavioral biometric data may be obviously utilized for one purpose or another. For example, in an application for biometric authentication, an evident measure of utility for the provider is its ability to verify the identity of an individual. Likewise, in an application based on human computer-interaction, the provider may require the behavior to still work as reliable input modality for computer systems. In a healthcare application the service provider may be interested in detecting abnormal behavior patterns, monitoring specific aspects of the behavior such as counting steps or inferring the  preferences of a user for personalization. The utility of the provided service may be assessed as the performance in carrying out those tasks.

\subsection{Privacy Concerns}
There are also troubling privacy implications derived from the
significant amount of personal information implicitly collected in behavioral data driven applications. 
 As we have seen, behavioral data can be used as biometrics because they are rich in individuating information. The counterpart is that any entity that collects behavioral data could use it to identify people even if that is not the main purpose of the service they provide. What aggravates this problem is that people might not be aware that they are being measured, either because of the lack of transparency and adequate consent frameworks, or because the surveillance is meant to be covert. But besides identity, behavioral data carry a wealth of potentially sensitive information that can also be abused. For example, behavioral traits like our voice, eye gaze, gait, or brain responses, are correlated with different diseases~\cite{yang2013specific, de2017freezing}, mental states and emotions~\cite{Sur2009, voice_emotion_recognition}, and specific involuntary reactions (such as pupil dilation) can signal our interests \cite{kroger2019does}.
 
\begin{figure}[!ht]
    \centering
      \includegraphics[width=0.49\textwidth]{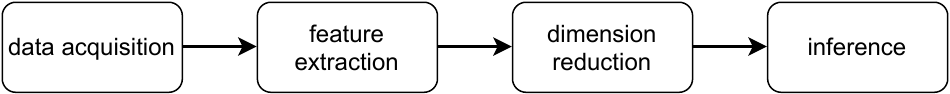}
\caption{The general behavioral-based inference process.}
\label{fig:biometric-recognition} 
\end{figure}

 Technically, the general process for inferring identity or other information about an individual from their behavioral data follows four steps, depicted in Figure~\ref{fig:biometric-recognition}. First, there is a data acquisition step in which the behavioral data are recorded and digitised. Then a feature representation that is suitable for the latter inference is extracted from the raw data. This feature representation is then usually reduced to lower the number of dimensions. In the last step the reduced feature representation is used to perform the inference of either identity or specific attributes. Thus, machine learning techniques are applied to classify the user data as belonging to an existing user profile or not, or as belonging to a specific attribute class (man, woman). Regression models can also be applied to assign the target individual with a measure (e.g., degree of depression on a continuous 1–5 scale). Based on this general workflow, a service that uses a voice-controlled personal assistant could apply the process to classify the user commanding to open an email application as the owner of the account (authentication). But it could also exploit the voice features to classify the mood of the user and offer them highly targeted advertisements, a practice that often comes with discrimination and threatens user's autonomy. Amazon, for instance, has a patent on technology to extract emotions from user's voice \cite{amazonPatent}.

%\iflong
%\else\fi

While big companies already collect a huge amount of behavioral data, the advent of affordable consumer wearables with numerous sensors (e.g., VR/AR devices with eyetracking, head pose detection, and electroencephalograpy (EEG) sensors) exacerbates the issue. Once the data are collected, even if for a legitimate, user-consented functionality like fraud detection based on behavior anomaly, these data can be exploited to learn private information. Hence, the need for techniques to protect behavioral data is poignant. To establish a map of current research on the topic, we categorize and analyze the existing
of protection approaches to prevent from identity and attribute disclosure.

\subsection{Attacker Model}
Our adversary is a malicious service or application provider that wishes to infer private information about the user. As the service provider the adversary has full access to the behavioral biometric data and can freely select an inference technique. Further, they also might have access to additional prior knowledge about the user such as biometric templates or soft biometrics.

%% file: sections/taxonomy.tex
Based on our literature analysis, we identify two main \textbf{privacy threats} that apply to behavioral data collected/processed by a third party and can be explained in terms of the related attacker model: 
 \begin{itemize}
      \item \textbf{Identity Disclosure.} The attacker's goal is to use the behavioral data to identify the user. In this threat model, we assume that the attacker is able to link the target's behavioral data to the target’s identity and now wants to identify them in another scenario. For example, linking the user account and data in a work-related application to their account in an entertainment application. This linkage would allow the attacker to learn more about the user activity. An example of this type of attacker, as presented in~\cite{Steil19ETRA}, could be a VR company with devices that record eye-tracking offering several services (e.g., games, adult content, professional training apps). This company would be able to determine if a user is the same person across these applications using their eye-tracking data, even if the user takes care to create accounts with different names or fake personal data. Moreover, it is not uncommon that behavioral data are sold to third parties or released unintentionally through a breach or hack\footnote{https://www.zdnet.com/article/over-60-million-records-exposed-in-wearable-fitness-tracking-data-breach-via-unsecured-database/}.
  
      \item \textbf{Attribute Disclosure.} In this threat model, the attacker goal is not to re-identify the user across accounts, but to derive sensitive attributes included within the available behavioral data that the user did not intend to disclose, such as gender, age, or mental state. The attacker might have had previous access or could have collected a dataset where to train the machine learning model for the targeted inference. For example, based on publicly available electroencephalogram datasets of alcoholic and non-alcoholic persons \cite{EEGDatabase, karamzadeh2015relative}, it could be possible to build a classifier that determines if newly gathered data from a entertainment application using a brain-computer interface (BCI) belong to a user with an alcohol problem.
 \end{itemize}

\iflong
\begin{figure}[!ht]
    \centering
      \includegraphics[width=0.30\textwidth]{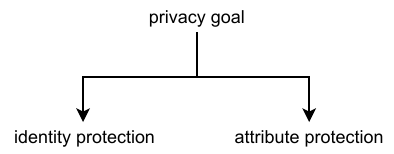}
\caption{Taxonomy of anonymization  techniques for behavioral data protection according to the privacy goal.}
\label{fig:taxonomy-goal} 
\end{figure}
\else\fi

From the privacy threats, we can derive the two \textbf{anonymization goals} with which techniques can be categorized, i.e., focused on protecting user \textbf{identity} and focused on protecting specific \textbf{attributes}\iflong, as depicted in Figure~\ref{fig:taxonomy-goal}\else\fi. 

\begin{itemize}
    \item \textbf{Identity Protection}. The process of transforming the behavioral biometric data of person in such a way that the person can no longer be linked to the data. \textbf{Pseudonymization} replaces the identity of a person with a new one and \textbf{anonymization} removes the identity altogether. 
    \item \textbf{Attribute Protection}. The process of transforming the behavioral biometric data of a person in such a way that specific private attributes of the person can no longer be inferred from the data. This encompasses both long-living attributes such as age or gender and short-living attributes such as mental state or temporary health conditions. An extreme version of attribute protection is template protection. For \textbf{template protection} the identity verification of the person, in the context of an authentication system, should be still possible while all attributes are protected. \del{Further, multiple templates of the same person must not be linkable to each other.}
\end{itemize}

%\iflong
\begin{figure}[!ht]
    \centering
      \includegraphics[width=0.49\textwidth]{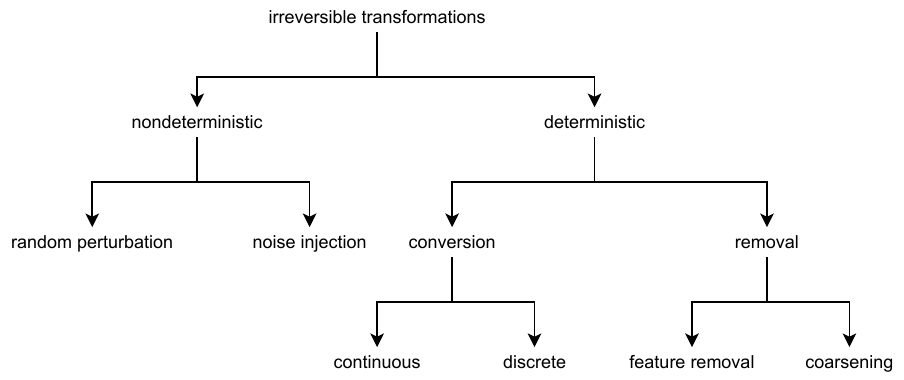}
\caption{Taxonomy of anonymization techniques for behavioral data protection according to the type of data transformation applied.}
\label{fig:taxonomy-transformations} 
\end{figure}
%\else\fi

We taxonomize anonymization solutions for behavioral biometric data according to the \textbf{type of transformation} applied, as depicted in Figure \ref{fig:taxonomy-transformations}. We include only fundemental concepts, some of the anonymization techniques combine multiple of them.
The basic and shared characteristic of all anonymization methods is that they are irreversible transformations, i.e. they cannot be reversed. The first distinction of our taxonomy is if they are deterministic or randomized techniques. \textbf{Non-Deterministic methods} rely on randomness in their transformation, which can yield different results for the same input and  \textbf{deterministic methods} always give the same result. There are several methods under these two approaches, as we detail in the following.

\begin{itemize}
\item \textbf{Non-Deterministic methods.}
\begin{itemize}
\item \textbf{Random perturbations.} A random transformation into a different domain.

\item \textbf{Noise injection.} Methods that add random noise to the data points.
    
\end{itemize}
\item \textbf{Deterministic methods.} Are further split into \textbf{removal} and \textbf{conversion}. The removal method eliminates data points from the data such that the data points do not have an influence on the anonymized result. Conversion methods transform the data points into a new representation, which typically depends on the original domain.

    \begin{itemize}
    \item \textbf{Removal.} Can happen in two forms: \textbf{\coarse} and \textbf{\feature}. Coarsening refers to removing parts of each data point or  making the data more sparse. Feature removal refers to removing data points belonging to a specific feature altogether.
    \item \textbf{Conversion.} Can be \textbf{discrete} or \textbf{continuous}, depending on if the result of the conversion is a discrete or continuous value.

\end{itemize}
\end{itemize}

%% file: sections/voice.tex
 %For the remainder of this section, we assume that the voice has been recorded via a microphone and therefore is no longer a continuous frequency spectrum but a discrete one.

%The human voice is the sound that is produced by the human sound organ and is unique for each individual. The voice is used during verbal communication which is called speech and is one of the main communication mediums between humans.

Voice processing and analysis~\cite{speech_processing_baeckstroem} have long been performed and hence a large set of specific terminology exists to describe it.\del{ A sound is a change in air pressure, which is often described as airwaves.} The sound of the human voice is created by the Larynx and then travels via the vocal tract, which transforms and filters the sound before it leaves the mouth. Due to its approximate tube shape, the vocal tract produces resonances of the sound which are dependent on the length of the vocal tract.
\del{Human \textbf{speech} is a sequence of sounds that convey meaning.} A Phoneme is the smallest unit of sound that distinguishes one word from another and an utterance is a unit of speech between two clear pauses.\del{ The frequency spectrum is the range in which the frequency of a sound signal can vary, it is gained from the original signal by using a fast Fourier transform (FFT). An important variant of the spectrum is the log-spectrum which allows a better human interpretation of a signal because the human perception of the magnitude of the signal is roughly approximated by the log transformation. Connecting the peaks in the log-spectrum gives the formant frequencies which correspond to the resonances in the vocal tract and uniquely identify vowels.} \new{The log-spectrum is an important representation of sound as it is closer to human perception.} By using a domain transformation (fast Fourier transform (FFT) or cosine) on the log-spectrum we get the cepstrum\iflong (see Figure~\ref{fig:speech_processing})\else\fi. The cepstrum is useful because it allows easy estimation of the fundamental frequency (f0) of the signal. The perceived fundamental frequency by humans is known as pitch. A widely used scale to transform the fundamental frequency to the pitch is the Mel scale. Using the Mel scale the cepstrum can be sampled at frequencies with the same perceived distance using weighted sums. Applying an FFT on those sums gives the Mel-frequency cepstral coefficients (MFCC). The MFCCs are an approximate quantification of the signal spectrum that focuses on the macrostructure of the signal.

\iflong
\begin{figure}[!ht]
    \centering
      \includegraphics[width=0.45\textwidth]{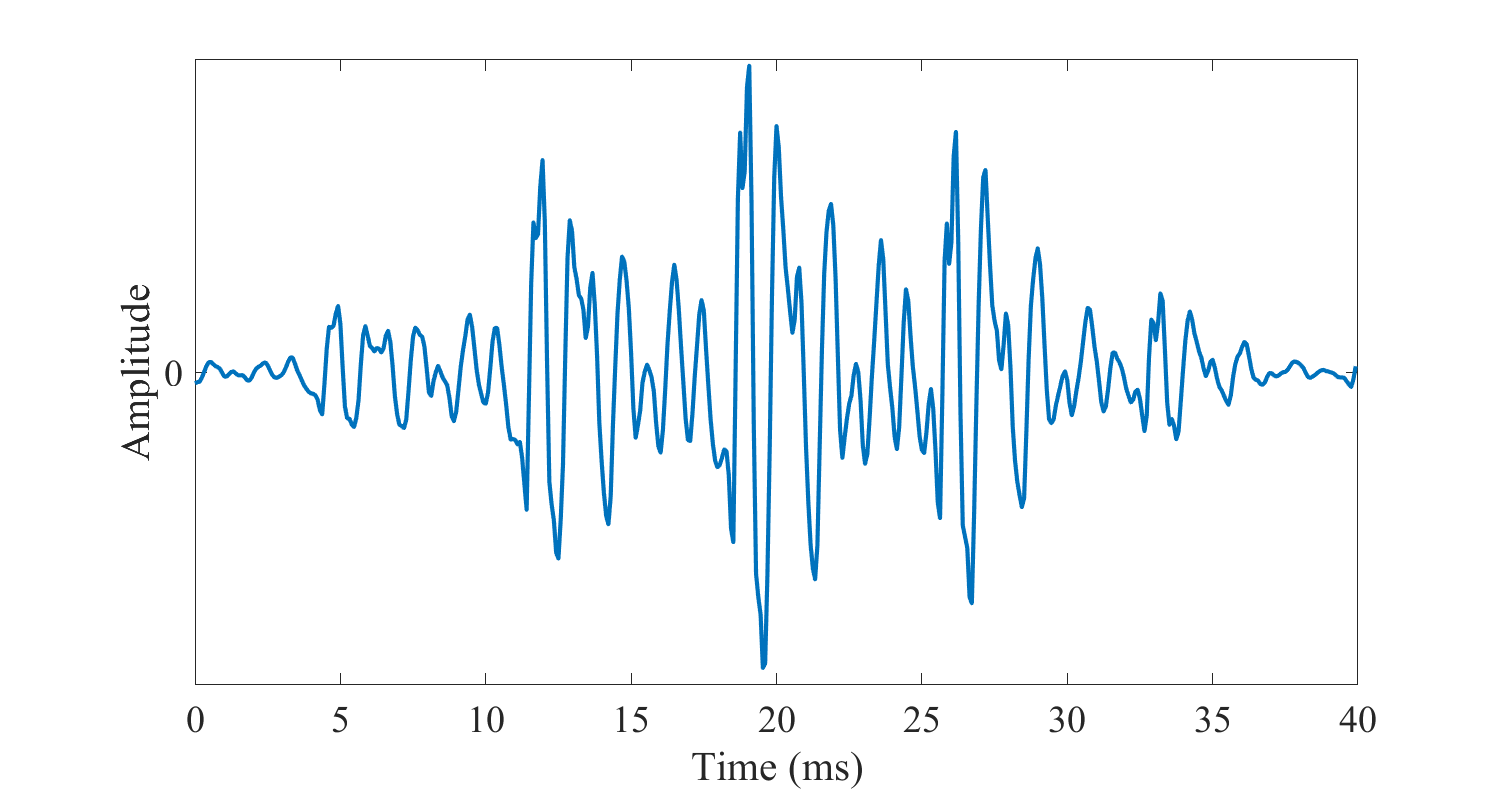}
       \includegraphics[width=0.45\textwidth]{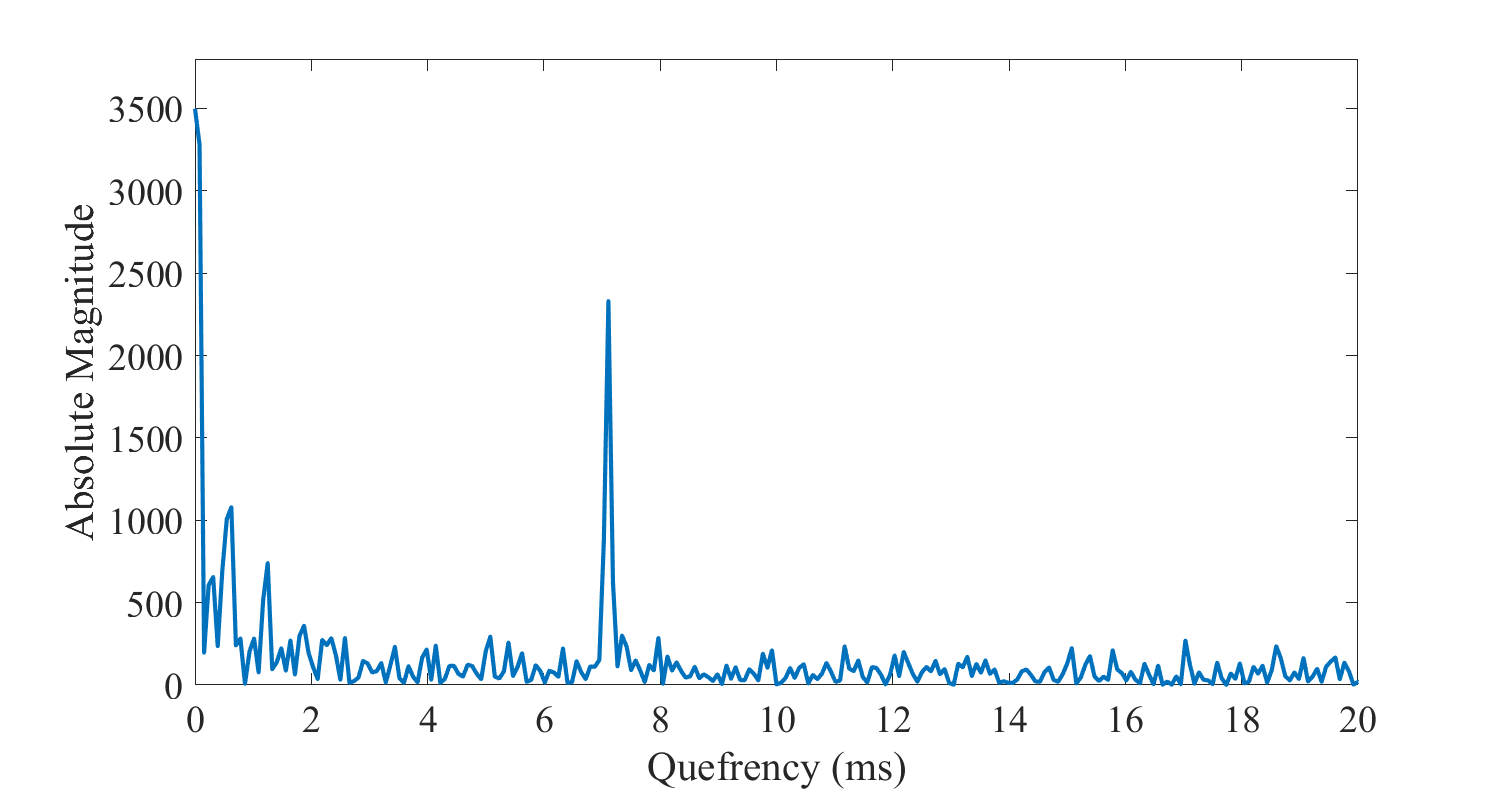}
\caption{A windowed speech segment (left) and its corresponding Cepstrum (right), Source: \url{https://wiki.aalto.fi/display/ITSP/Cepstrum+and+MFCC}.}\label{fig:speech_processing}
\end{figure}
\else\fi

The following gives a short overview of the field of speaker recognition which aims to establish the identity of a speaker. Gaussian mixture models~\cite{reynolds_speaker_1995} (GMM) represent speakers as the distribution of their feature vectors. The feature vectors are extracted from the speech (most often represented as MFCC) of the speaker and then modeled as Gaussian mixture density. A GMM assumes that the data points are generated by a finite number of Gaussian distributions with unknown parameters. Each feature vector is represented as a linear combination of Gaussian densities. 
A universal background model (UBM) is a GMM that models a wide variety of non-target speakers, representing possible imposters. The means of the UBM are then adjusted to the target speaker by using a maximum aposteriori adaption~\cite{reynolds_speaker_2000} resulting in a GMM for the target speaker. The benefit of this approach is that the Gaussians used to model the target speaker are the same as in the UBM. For the classification of a speaker, the log-likelihood of the target speaker GMM is compared to that of the UBM to determine if the speaker should be accepted. An alternative to the log-likelihood approach is to get a GMM for each speaker recording through a maximum a posteriori probability (MAP) adaptation of the UBM and then map these GMM to a new feature vector, called Supervector~\cite{campbell_support_2006}. Supervectors can be classified using traditional methods like support vector machines. A common extension of Supervectors is the total variability (TV)~\cite{dehak_front-end_2011} approach. This maps the Supervectors to a low-dimensional space that models both the speaker and the channel variability. The resulting vector is called i-vector and is the de facto state-of-art in speaker identification. An alternative to i-vectors are x-vectors~\cite{snyder_x-vectors_2018} which are extracted for each utterance via a deep neural network (DNN).

%Its name is a reflection on the circumstance that the cepstrum is a rearrangement of the original signal.

\subsubsection{Utility}
The main usage of voice recordings is the transmission of information between humans, however, in recent years voice also became an important input modality for computer systems~\cite{povey2011kaldi}. In both cases, it is important that the content of the speech is intelligible for the intended listeners. But also the mere detection of speech in audio samples can be useful, for example for crowd detection~\cite{cohen-hadria_voice_2019}. Further, voices uniquely identify their speaker, making them suitable both for authentication and recognition purposes~\cite{rui2018survey}.

\subsubsection{Threat Space}
The privacy threats for human voices range from the identification of individuals, over the inference of private attributes, to identity theft via fake recordings. The identification of individuals via their voice has long been apparent to humans. But voices convey more information than just identity, they also allow us to infer attributes such as gender~\cite{voice_gender_recognition}, or emotional state~\cite{voice_emotion_recognition}. Further, modern speech synthesis methods allow the creation of fake voice recordings for a target speaker, enabling identity theft or the circumvention of speaker authentication systems. Other than the other behavioral biometric traits voice and its resulting speech can also carry a semantic meaning, which can be privacy sensitive.

\subsubsection{Privacy Goals}
Voice has speech blurring as an additional privacy goal, which aims at destroying the intelligibility of the speech to protect its semantic content from unintended listener.

\subsubsection{Anonymization Techniques}
We now present the surveyed anoymization techniques that deal with protecting human voices.

\paragraph{Random Perturbation}
Parthasarathi et al.~\cite{parthasarathi_wordless_2013} extend their feature removal methods~\cite{parthasarathi_lp_nodate} by additionally shuffling the voice blocks adding randomness. Mtibaa et al.~\cite{mtibaa_cancelable_2018} propose a template protection scheme that relies on shuffling the feature vector of a GMM-UBM speaker identification system.

\paragraph{Noise Injection}
Tamesue et al.~\cite{tamesue_sound_2014} propose a very simple method to make speech unintelligible by simply playing pink noise between 180 and 5630 Hz with various dBs.
\new{
Ma et al.~\cite{ma_you_2021} also try to make speech unintelligible but focus on smartphone recordings. Their device creates two ultrasound waves whose interaction creates low frequency waves that noise the microphone of a smartphone but cannot be heard by humans. The waves are generated in a random fashion. In their evaluation they found that they can block smartphone recordings up to 5 meters, depending on the type of smartphone.
}

Hashimoto et al.~\cite{hashimoto_privacy-preserving_2016} proposes a system to preserve speaker privacy in physical spaces. The core idea is to add white noise to prevent recordings of speakers to be used for identity theft.\del{ They found that increasing the Signal-to-noise ratio (SNR) is bad for the intelligibility of the speech and experiment with filtering the white noise in frequency ranges from 0 to 8 kHz to boost the performance of the scheme.} They conclude that preventing speaker identification is possible (equal error rate (EER) from 2\% to 17\%) while at the same time keeping the intelligibility of the speech at a high level (short-time objective intelligibility~\cite{STOI} from 1 to 0.9).

Ohshio et al.~\cite{ohshio_active_2018} train multiple so-called babble maskers from pre-recorded speakers by segmenting the speech and then averaging the segments. When a speaker should be de-identified the babble masker is selected based on the fundamental frequency and the pitch of the person.
Vaidya et al.~\cite{vaidya_you_2019} proposes to add random noise to four features: pitch, tempo, pause, and MFCC. We found the descriptions of their approach to be rather short.

Two methods have been proposed that rely on differential privacy for noise injection.
Hamm et al.~\cite{hamm_enhancing_2017} proposes a differential private min-max filter. The min-max filter minimizes the privacy risk while maximizing utility risk with a given utility and private task. The differential privacy is achieved by adding noise either in front of the filter or after the filter.
Han et al.~\cite{han_voice-indistinguishability_2020} rely on X-vectors as speaker representation and formally define voice-indistinguishably a privacy metric using differential privacy. As a measurement of similarity between x-vectors the angular distance is used and the overall scheme gives an upper limit of this distance until which two x-vectors cannot be distinguished. We note that a discussion of sensitivity, which is required for differential private mechanisms, is missing from this work.

\paragraph{Feature Removal}
%Feature removal is the process of removing features that can be used to perform inferences.
Parthasarathi et al.~\cite{parthasarathi_speaker_2009} propose three feature removal methods for privacy-aware speaker change detection. Adaptive filtering assumes that the excitation source is independent of the vocal tract response. They perform short-term linear prediction analysis to estimate an all-pole model~\cite{all-pole-model}(representing the vocal tract), a residual (representing the excitation source), and the gain. Then the residual is used to estimate its real cepstrum. 
Their second method is to remove all subbands except the one from 1.5 kHz to 2.5  kHz and from 3.5 kHz to 4.5 kHz. They represent the two subbands as MFCC coefficients and log-energy from a single filter. Their last method only uses the spectral slope of the speaker represented as cepstral coefficients. In another work~\cite{parthasarathi_lp_nodate} Parthasarathi et al. also propose similar feature removal methods for speaker diarisation using the real cepstrum and MFCC as features. Their analysis finds that MFCC works better than real cepstrum. Additionally, they add subband frequency information between 2.5 kHz and 3.5 kHz and the spectral slope. The privacy is evaluated by trying to recognize phonemes in the anonymized speech using an hidden Markov model (HMM) GMM speaker diarisation method as an evaluation system.

Wyatt et al.~\cite{wyatt_conversation_nodate} propose a feature removal method for speaker segmentation and conversation detection. They split the audio into segments and save for each the non-initial maximum autocorrelation peak, the total number of autocorrelation peaks, the relative spectral entropy, and the energy of the frame. Zhang et al.~\cite{zhang_privacy-preserving_2012} uses the same features as proposes by Wyatt et al. except for the energy of the frame and then use an HMM to perform the conversation detection. An evaluation of privacy is missing in both works.

\new{
Ditthapron et al.~\cite{ditthapron_privacy-preserving_2021} have investigated how speech from non-target speakers can be removed in a speech assessment scenario. To separate the speakers they first extract speaker representations from the MFCC of the speech via an encoder. The speaker representation is then concatenated with the original MFCC before all but the target speaker are filtered out in the speaker matching network. We are missing a convincing evaluation of privacy.
}

\del{
In~\cite{nelus_gender_2018} and~\cite{nelus_privacy-aware_2019} Nelus et al. propose to use a DNN to extract features from a speaker that allow gender recognition but not speaker identification.}
Nelus et al.~\cite{nelus_gender_2018} propose to train a DNN via adversarial learning to extract features from a speaker that allow gender recognition but not speaker identification. Their evaluation shows a drop in identification from 61\% to 26\% while the gender recognition only drops by 1\%. They also proposed a similar system~\cite{nelus_privacy-preserving_2021} which remove speaker identities from urban sound recordings.
Cohen-Hadria et al.~\cite{cohen-hadria_voice_2019} also use a neural network and use it to extract the voices from recordings that consist of both background and voice noise in which the voices should be anonymized. They remove attributes with two methods. The first method simply low-pass filters the voice at 250 Hz. The second method extracts the MFCC from the voice and then uses the first 5 components to create a new voice. In the end, the blurred speech is recombined with the background noise. Evaluating with a speaker identification system they were able to reduce the identification down to 29\% from 43\%.

\paragraph{Discrete Conversion}
%Template protection deals with the protection of biometric templates for authentication. Its two main goals are to protect all attributes of the speaker and to prevent the linkability of multiple templates of the same person.

For discrete conversions we found multiple template protection schemes.

Pathak et al.~\cite{pathak_privacy-preserving_2012} present a hashing algorithm to protect voice data for authentication purposes. The supervector of a speaker is gained by performing the MAP adaptation of a universal background model for each utterance of the speaker and concatenating the means of the adapted model.\del{ This supervector is the feature vector for the classification.} The locality sensitive hashing is then performed with the supervector which transforms it into a low dimensional space, which is referred to as a bucket. This operation is an approximation of the nearest neighbors algorithm allowing the comparison of buckets to authenticate the individual.\del{ In order to make the representation more privacy preserving the salted hash of the result is computed.} 

Portelo et al.~\cite{portelo_secure_nodate,portelo_privacy-preserving_2014} propose a template protection scheme based on secure binary embeddings. The authors use a speaker identification system that uses supervectors and i-vectors to represent the features of a speaker's voice. The feature vectors are then encoded with secure binary embeddings which have the property that if the euclidian distance of the two vectors is below a certain threshold then the hamming distance of the resulting hashes is proportional to the euclidian distance. This allows the comparison of the encoded vectors by using a support-vector machine (SVM) with a hamming distance-based kernel. Billeb et al.~\cite{billeb_biometric_2015} propose a template protection scheme that is based on fuzzy commitment. They first extract the frequency spectrum via an FFT and then extract features from the magnitude spectrum. Then the MAP adaptation of a GMM-UBM speaker identification system is applied and additional statistics are extracted. The template is then stored as a combination of error-correcting code and hash algorithm.

\paragraph{Continuous Conversion}

\textbf{Speaker transformation} is the process of manipulating the voice characteristics of a speaker (not the linguistic features) to make the voice sound like a target speaker. A target speaker can be either a specific natural speaker or a synthetic speaker. For the synthetic speaker either an existing speaker is used or a new one is generated, for example by averaging multiple speakers into one. The general approach of speaker transformation is that the voice characteristics of the source speaker are extracted and then transformed to match the target speaker. In the last step, the new speaker is synthesized. The following methods perform speaker transformation.

Jin et al.~\cite{jin_voice_2009} evaluate four methods for speaker transformation for identity protection. Their base method uses a GMM-mapping based speaker transformation system to transfer speakers to a target synthetic voice called kal-diphone. Further, they test duration transformation in which the length of utterances of the source speaker is scaled to match the ones of the target speaker. Double voice transformation simply repeats the process of mapping the source to the target twice(8\% identification accuracy). Lastly, they try an extrapolated transformation in which they use the linear mapping of the source to the target to extrapolate beyond the target (0\% identification accuracy). Pobar et al.~\cite{pobar_online_2014} also use a speaker transformation system based on GMM mapping but combine it with a harmonic stochastic model. The system is trained on a set of speakers to learn the transformation functions. Instead of retraining the system for a new speaker one of the existing transformation functions is applied. This removes the need for a parallel corpus for the speakers that should be protected. The target speaker is a synthetic speaker which reduces the identification accuracy from 97\% down to 9\%. Justin et al.~\cite{justin_intelligibility_14, justin_speaker_2015} investigate the intelligibility of transformed speakers. They test with a diphone speech synthesis system and an HMM-based speech synthesis system to transform speakers into a synthetic speaker. They performed a survey with human listeners to evaluate the intelligibility of the protected speakers, measuring the word error rate. Abou-Zleikha et al.~\cite{abou-zleikha_discriminative_2015} do not propose a speaker transformation method themselves but explore how to select a target speaker to achieve the lowest identification rate and have good results when the speaker is transformed back to the source speaker. They formulate this as an optimization problem and measure the distance between two speakers with a confusion factor, for which they evaluate entropy and Gini index as metrics. Pribil et al.~\cite{pribil_evaluation_2018} propose a speaker de-identification method that relies on modifying several features of the source speaker. In the first step, the prosodic and spectral features are extracted from the source speaker. They then modify the features to make the speaker sound older, younger, more female, and more male by using manually defined transformation functions and feature differences for each class. After the features are modified the de-identified speaker is synthesized.

Bahamanienezhad et al.~\cite{bahmaninezhad_convolutional_2018} have developed a speaker transformation method that uses a convolutional encoder/decoder network. They, first extract spectral features and excitation features (f0) from the source speaker. The spectral features are then mapped via the encoder/decoder framework to a target speaker. The resulting speech is fused together either via taking the average or via a gender-based average to create an average speaker. From the excitation features, only the fundamental frequency is transformed via linear transformation, the remaining features stay the same. Both spectral and excitation features are used to synthesize the de-identified speaker.
Fang et al.~\cite{fang_speaker_2019} use a similar averaging approach but rely on x-vectors. They extract the x-vector of a speaker and then use a set of random x-vectors of unrelated speakers to calculate a mean x-vector. They also propose to construct an altogether new x-vector that has a similarity scoring of s to the original x-vector. Further, they keep the fundamental frequency of the speaker the same. In their evaluation they demonstrate EER up to 34\% for their anonymization. \new{Mawalim et al.~\cite{mawalim_speaker_2022} propose to improve the system by Fang et al. by scaling the f0 frequency either up or down, increasing the length of the speech utterances by 1.2, and using singular value modification for the combination of the x-vectors. Their EER improved up to 54\%.  Further improved was this system by Prajapati et al.~\cite{prajapati_voice_2021} who added a CycleGAN to modify the speakers.}

Kesking et al.~\cite{keskin_measuring_2019} do not study de-identification directly but instead try to create an imposter transformation for a target speaker. They use a cycle generative adversarial networks (GAN) voice converter to transform speakers and then evaluate against four speaker identification systems to see if the target speaker is recognized.\\

%End speaker transformation.
\textbf{Frequency warping} is a technique that is similar to speaker transformation, the main difference is that frequency warping focuses on transforming the frequency spectrum of a speaker and usually does not try to transform the source into a specific target speaker. It is mostly used for identity and gender protection. A common goal of frequency warping is vocal tract length normalization in which the resonances that are specific to an individual's vocal tract length should be removed or altered.

Faundez-Zanuy et al.~\cite{faundez-zanuy_speaker_2015} explore two approaches for gender protection: Phase vocoder and vocal tract length normalization. The vocoder approach detects peaks in the voice signal. For each peak, a bin is defined and compared to its two neighbors to define a region of influence. Then the peak and its region of influence are shifted by a peak specific frequency. In the last step artifacts from the shift are removed. The vocal tract length normalization approach defines frames on the signal spectrum and stretches or compresses them using a frequency warping function. For both genders they can reduce gender recognition to chance level, however the identity recognition is also close to chance level.

Valdivielso et al.~\cite{abad_advances_2016} present a speaker protection approach that transforms the pitch and the frequency axis. Further, the parameters of the transformation are embedded into the signal for later re-identification. Lopez-Otero et al.~\cite{lopezotero_influence_2017} rely on frequency warping and amplitude scaling for speaker protection in the context of depression detection. They implement both operations as an affine transformation in the cepstral domain and manually define piece-wise linear transformation functions. They demonstrate an increase of the EER from 9.7\% to up to 44\% for the speaker identification, while the depression detection stays similar to the clear data. 

Magarinos et al.~\cite{magarinos_reversible_2017} also rely on frequency and amplitude warping for speaker protection. First, they extract the cepstral voice vectors from the speaker and then convert them into a discrete spectrum. Then dynamic frequency warping (DFW) is applied to map the source spectrum bins to the target spectrum. As multiple source bins can have the same target bin, all source bins that map to the same target bin are averaged. Additionally to the frequency and amplitude warping the fundamental frequency is adjusted regarding its mean and variance. They demonstrate an identification reduction from 99\% to 4\%. 
Aloufi et al.~\cite{aloufi_emotionless_2019} try to hide the emotional state of speakers before their speech is sent to a voice-based cloud service. They first extract the fundamental frequency, spectral envelope, and aperiodicity. The features are then transformed via a CycleGAN from emotional speech to neutral speech. \new{ In a second paper~\cite{aloufi_privacy-preserving_2020} the same authors propose a full framework for anonymization which uses an variational autoencoder. Their framework has three modi, the first removes private attributes, the second removes the identity, and the third removes the intelligibility of the speech. Specific to this approach is that two separate encoders are used, one to encode the speech and one to encode the speaker. Their results for hidings the emotional state show a reduction from over 70\% to about 20\% and for hidings sex a reduction from up to 99\% to the chance level of 50\%}

Srivastava et al.~\cite{srivastava_evaluating_2020} evaluate multiple speaker protection methods against an informed attacker. They work with three attacker models: An ignorant attack that is not aware that the voice data is de-identified, a semi-informed attacker that knows that the data is de-identified, and an informed attacker that knows the de-identification method and its parameters. The first method is a vocal tract length normalization approach. The speaker is represented as a set of centroid spectra. The algorithm then calculates the closest path between the source set and the target set to get the parameters for the warping. 
The second method uses a neural net encoder/decoder approach to transform the speaker. They found large differences for the different attacker models, while the ignorant attacker can achieve EER of up to 50\% the informed attacker only achieves 11\% as its highest EER. This finding highlights how important strong attacker models are for the evaluation of anonymization techniques.

\new{
Ali et al.~\cite{ali_privacy_2021} also propose an autoencoder to anonymize at the network edge specifically for the input of voice assistants. Their idea is to extract privacy friendly features by training classifiers on the latent code of the voice samples. They use the trained classifier to perform gradient reversal on the encoder to unlearn the features learned for identity, gender, and language.
}
\new{Yoo et al.~\cite{yoo_speaker_2020} use a CycleGAN for speaker anonymization which uses a variational autoencoder as its generator. They train against a DNN speaker recognition system as the discriminator.\\
}
\new{
Patino et al.~\cite{patino_speaker_2021} pseudonymize speakers by transforming their McAdam coefficients. In the first step linear predictive coding (LPC) is applied to an input speech frame. The coefficients of the LPC are then transformed into poles and the poles which have a non zero imaginary part are shifted according to the angle between the real and imaginary part of the pole. Then the resulting poles are transformed back into speech. Their evaluation shows that this approach performs well against an ignorant attack which is not aware of the anonymization increasing ERR from 3\% to 26\% while an informed attacker still achieves 5\% ERR. Gupta et al.~\cite{gupta_design_nodate} further improve on transforming the McAdams coefficients by not only changing the angle of the complex poles but also modifying their radius.
}

%End Frequency warping
\paragraph{Continuous Conversion + Random Perturbation}
Canuto et al.~\cite{canuto_effective_2014} proposes a new method for template protection in which the feature vector is shuffled via a randomized sum. For each feature vector, the elements are shuffled based on a secret key. Two random vectors of the same length are derived from the key. These vectors give the position of the attributes that should be summed. The reorganized feature vector is summed up with the vectors resulting when the position vectors are applied to the original feature vector.

\paragraph{Continuous Conversion + Noise Injection}
Kondo et al.~\cite{kondo_towards_2013,kondo_gender-dependent_2014} create so-called babble maskers by segmenting speech into ten second segments and then averaging them into babble maskers. Besides speaker-dependent maskers, they also create gender-based babble maskers based on multiple speakers of the same gender. The babble masker is then applied to the recording of the speaker.
Qian et al.~\cite{qian_hidebehind_2018} present a method to sanitize speech before it is sent to the server of a virtual assistant. Their main method is to perform vocal tract length normalization via a compound frequency warping function consisting of a bilinear and a quadratic function to avoid re-identification attacks. The parameters of the warping function are selected randomly. Additionally, they add Laplace noise after the warping function to make the anonymization more robust. For the result, they claim to achieve differential privacy. In a follow up work~\cite{qian_speech_2021} the same authors further investigate the security of their scheme. Srivastava et al.~\cite{srivastava_evaluating_2020} also investigate the security of the scheme with stronger attackers.

\subsubsection{Evaluations}

Most of the reviewed works evaluate the quality of the de-identification by comparing the recognition rates of attributes or identities on unmodified and de-identified data. The recognition is done via machine learning models or human listeners. As metrics to measure the recognition rate the papers mostly rely on the equal error rate (EER), false positive rate (FPR), false negative rate (FNR), recall, precision, and F1 score. Abou-Zleikha et al.~\cite{abou-zleikha_discriminative_2015} also use entropy and the Gini index to evaluate the de-identification performance.
Additionally to the de-identification, some works evaluate the loss of utility. One important goal in regards to human listeners is to achieve a natural-sounding de-identified voice. The naturalness is evaluated by human listeners using the mean opinion score. Another important aspect is the intelligibility of the de-identified speech. Intelligibility can be evaluated via human listeners or machine learning models using the word error rate, phoneme error rate, or short-time objective intelligibility.
A common limitation we observed is that most evaluations use the clear data to train the recognition model and then test it against the anonymized data. This approach implicitly assumes that the attacker is not aware of the anonymization and hence does not try to circumvent it. Srivastava et al.~\cite{srivastava_evaluating_2020} explicitly assume an attack on the anonymization proposing attackers with varying degrees of information about the performed anonymization.

\new{The VoicePrivacy challenge~\cite{tomashenko2022voiceprivacy} is an initiative to improve the evaluation methodology in the field of speaker anonymization. They use EER and the log-likelihood-ratio cast function (Cllr) to evaluate speaker verifiability and word error rate to evaluate speech intelligibility. In a post evaluation they also retrained their speaker verification systems with anonymized speech data to test against an informed attack. Their methodology is already being applied by others~\cite{kai_lightweight_2021} to compare speaker anonymization methods.}
Qian et al.~\cite{qian_towards_2018} present a framework to reason about the privacy and utility of voice anonymization techniques. For this, they present the measure of p-leak limit which should give a maximum privacy leakage per speaker for a published dataset. Zhang et al.~\cite{zhang_enhancing_2020} propose a theoretical framework to quantify the privacy leakage risk and utility loss for speech data publishing. For speaker de-identification they do not describe their own speaker de-identification techniques but give a framework for quantifying the utility privacy loss.

%% file: sections/gait.tex
The human gait is the pattern in which humans move their limbs during locomotion, multiple manners of gait exist such as trotting, walking, or running. Gait can be broken down into individual gait cycles~\cite{gait-cycle} \iflong(see Figure~\ref{img:GEI})\else\fi which is the shortest repetitive task during the gait. The gait cycle spans from a specific gait event of one foot until the same foot reaches the same gait event. It consists of a stance phase, in which the foot is on the ground, and a swing phase, in which the foot is in the air. The two phases alternate for each foot.
Due to its usefulness as a behavioral biometric trait for identifying individuals, gait has long been a research interest of both computer science and psychology. For example, Yovel et al.~\cite{yovel_recognizing_2016} find that it plays an important part for humans to identify people at a distance, and Pollick et al.~\cite{pollick_gender_2005} show that it is possible for humans to infer the gender of a walker, even when the walker is only shown as a set of points, as so-called point-light-display. The following section deals with the anonymization of gait patterns.

%https://www.protokinetics.com/understanding-phases-of-the-gait-cycle/

\iflong
\begin{figure}[!ht]
    \centering
      \includegraphics[width=0.49\textwidth]{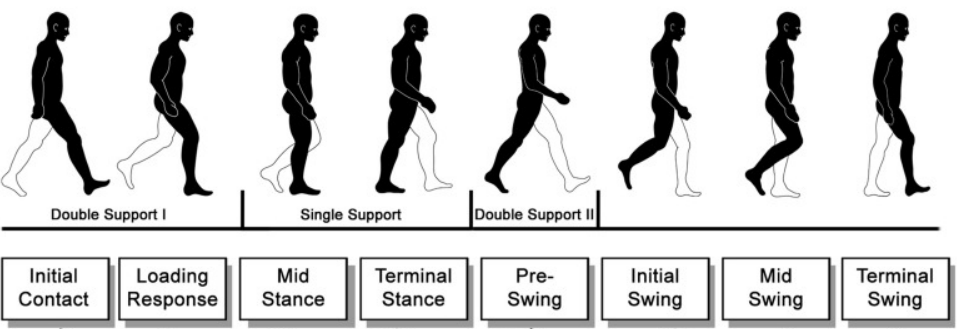}
\caption{The phases of the gait cycle, source:~\cite{gait-cycle}.}\label{img:GEI}
\end{figure}
\else\fi

Gait recognition methods have been an active research topic in the past, hence a large set of different methods for various capture methods exists. Wan et al.~\cite{wan2018survey} performed a recent survey on the subject and list recognition methods for cameras, accelerometers, floor sensors, and radars.
The main portion of the works focuses on camera based gait recognition which is classified by Wan et al. as either model-based or model-free. Model-based methods use a specific model of the walker, for example, a pendulum model of the legs, to then match the walker to it. Model-free methods, however, do not have an explicit model but rather use the entire capture of the gait to perform the recognition, for example by averaging the silhouette of the walker over time as a gait energy image. 
Accelerometer-based systems also average the gait into a feature representation either by segmenting the gait into its gait cycles or by using frames with a fixed size.

\subsubsection{Utility}
The human gait is omnipresent in everyday life and as such often captured as a byproduct of recordings being made. As such it is often not necessary to preserve the utility of the gait, but rather the utility of the recording. One example of this would be video recordings of people walking, the gait pattern itself is not so important but rather that the video looks natural and convincing to its viewers~\cite{ivasic-kos_person_2014}.
But there also exist use-cases in which the gait pattern itself should be captured, for example for medical examinations by a physician to find gait abnormalities~\cite{kirtley2006clinical}. Another more casual example would be the recording of the gait pattern to count the steps a person has performed during a day~\cite{7272064}.

\subsubsection{Threat space}
Due to its omnipresence in everyday life, human gait is easy to capture, especially because most capturing methods are unintrusive and do not require the participation of the victim. Additionally, it has been shown that gait recognition is very robust to video quality and obfuscation making it very much suited for surveillance systems~\cite{wan2018survey}. Besides identifying humans it has also been shown that gait can be used to infer private attributes like gender~\cite{pollick_gender_2005}. Considering all this the threat to gait biometrics is already large. What's more, with recent developments in richer capturing methods such as LiDAR~\cite{lidar_gait_recognition} or cheap motion capture suits, it is to be expected that the threat space for gait will even increase in the coming years.

%\subsubsection{Privacy goals}
%The privacy goals for gait do not differ from the general privacy goals.

%Gait recognition methods have been an active research goal in the past, hence a large set of different methods for various capture methods exists. Wan et. al.~\cite{wan_survey_2019} performed a recent survey on the subject and list methods for cameras, accelerometers, floor sensors, and radars.

%They only classify camera based systems this way.
%Further, they classify the methods by their feature representation as either model-based or model-free. Model-based method use a specific model of the walker, for example a pendulum model of the legs, to then match the walker to it. Model-free methods, however, do not have an explicit model but rather use the entire capture of the gait to perform the recognition, for example by matching the silhouette of the walker over time.

%For camera based systems the most prominent approach is model-free feature representation of using so called gait energy images. Gait energy images take the silhouette of the walker and then average it overtime into a single greyscale image. For accelerometer based systems the data is segmented in two different ways, either as extraction of the gait cycle, which then is average into a gait cycle template or as fixed size frames for which a feature vector is extracted.

%The classification of the feature representations is done with a classifier.
%\sh{Make a general section about the classification pipeline, Sensors recording, feature extraction/representation, dimension reduction, classification}

\subsubsection{Anonymization Techniques}

In the following, we present the gait anonymization methods found in the literature, sorted by our taxonomy.

\paragraph{Random Perturbation}
Hoang et al.~\cite{hoang_gait_2015} propose a fuzzy commitment scheme based on Bose–\linebreak[0]Chaudhuri–Hocquenghem (BCH) codes for storing accelerometer gait templates. After the feature extraction and binarization of the accelerometer data the reliable bits are extracted. These bits are then XORed with the BCH encoded secret key to gain the secure $\gamma$. Additionally to the $\gamma$, the hash of the secret key and some helper data are stored. During the authentication phase, the extracted reliable bits are XORed with the secure $\gamma$ and then decoded with BCH. The result can then be hashed and compared to the hash of the secret key.

\paragraph{Noise Injection}
The influence of noise injection on the performance of accelerometer/gyroscope authentication systems was studied by Matovu et al.~\cite{matovu_jekyll_2018}. For their approach, they generate a time series of noise values drawn from a uniform distribution and then merge the original time series with the generated one.\del{ The two traits evaluated are gait and handwriting.}

A noise injection approach for gait in videos was developed by 
Tieu et al.~\cite{tieu_approach_2017}. They use a convolutional neural network (CNN) to mix the gait of a second person (noise gait) into the original gait. In the first step, the silhouette for both the original and noise gait is extracted from a black and white representation of the input videos. The noise gait is selected hereby to have the same size and view angle as the original gait to achieve a more natural result. The silhouettes are then fed into the CNN which uses shared weights networks to abstract them and then merges the abstracted representations via a third network. In a post-processing step, the original gait is replaced with the newly merged gait. Depending on the view angle they achieve identification rates between 20\% and 1\%. 
The authors further improve their method in a follow up paper~\cite{tieu_spatio-temporal_2019}. Here the noise gait is generated via a generative adversarial network (GAN) that takes Gaussian noise as input and outputs noise silhouette. Instead of using a CNN they then use a self-growing and pruning GAN (SP-GAN) to fuse the noise and original gait. Here the identification accuracy was between 30\% and 10\%. Further, they propose an approach to colorize the resulting black and white silhouette\new{~\cite{tieu_color_2020}}. \del{ In ~\cite{tieu_rgb_2019} it is proposed to use a deep convolutional GAN to fuse original and noise gait which is further improved by transferring the colors without ground truth~\cite{tieu_color_2020}.}
\new{Hanisch et al.~\cite{Understanding_Hanisch_2023} investigated multiple anonymization techniques to protect identity and gender of walkers recorded via motion capture suits. One of their techniques was to add Laplace noise to all body positions of the walker, however their results show that effectively anonymizing was not possible without destroying the utility (measured as naturalness via a user study).}

\paragraph{Feature Removal}
A  feature removal approach for privacy-preserving activity recognition via accelerometers is proposed by Jourdan et al.~\cite{jourdan_toward_2018}. They extract various temporal and frequency features from the accelerometer data such as mean, correlation, energy, or entropy. Via experiments, they then determine the influence of each feature for activity and identity recognition. They find that the temporal features contribute more to identity recognition and frequency features more to activity recognition, therefore they remove the temporal features. Their results show a good trade-off between activity recognition (96\% reduced to 87\%) and identification (90\% reduced to 40\%). 
\new{Garofalo et al.~\cite{garofalo_siamese_2020} propose a temporal convolutional network as feature extractor which is trained via adversarial training. After the feature extractor created a feature vector it is evaluated by an identity verifyer and an attribute classifier which results are then used as the loss function for the feature extractor training.}
\new{Another technique tested by Hanisch et al.~\cite{Understanding_Hanisch_2023} was to remove body parts from gait motion capture data to see their impact on the recognition of identity and gender. They found that the gait data is very redundant and even when only the data for the head is kept identification is still close to 60\%}

\paragraph{Continuous Conversion}
A continuous conversion approach is blurring, in which persons in videos, including their gait, should be de-identified. As a first step, the silhouettes of the persons in the videos are tracked and segmented to then apply the blur. Agrawal et al.~\cite{agrawal_person_2011} proposed two blurring approaches exponential blur and line integral convolution (LIC). Exponential blur regards the video as a 3D space with the time as the z-axis and then calculates a weighted average of the neighbors of each voxel to blur via an exponential function. LIC works with the bounding box of the walker silhouette and maps it onto a vector field which is then used to calculate the output pixels. To counter reversal attacks against the blur randomization of the blurring functions at each pixel is proposed. Another blurring approach is proposed by Ivasic-Kos et al.~\cite{ivasic-kos_person_2014}. They apply a gaussian filter to blur the silhouettes of walkers. The filter calculates a weighted average of the color of the neighboring pixels, with the weights decreasing monotonically from the central pixel. \new{Thapar et al.~\cite{thapar_anonymizing_2021} consider the anonymization of gait in egocentric videos, which are videos that are recorded from a first-person perspective. They first learn the identities of gallery videos via the rotation of the camera which is then transformed into the camera rotation signature via guided backpropagation. This camera signature is then applied to the target video, mixing the gallery identity and the target identity. In their evaluation they test the identification of persons and find that the EER increases from around 20\% to around 50\% while the activity recognition is reduced by about 10\%}

\paragraph{Continuous Conversion + Discrete Conversion}
An approach that combines both continuous and discrete conversions for walkers in videos is proposed by Hirose et al.~\cite{hirose_anonymization_2019}. First, they extract the silhouette and the gait cycle of the walker. The silhouette is then transformed via a deconvolutional neural network encoder into a silhouette code. The code is converted by using a k-same approach in which the k-nearest neighbors of the input code are selected and then a weighted average is computed. The gait cycle is transformed via a continuous, differentiable, and monotonically increasing function. In the last step, the new video is generated by feeding the perturbed silhouette code and gait cycle into the convolutional neural network decoder. Their evolution shows that the gait recognition drops from about 100\% down to 29\%, 21\%, and 4\% depending on the recognition model.

\subsubsection{Evaluation}

Gait de-identification is evaluated in the literature via gait recognition systems or human observers with the recognition accuracy as the main metric, but there are also usages of the F1 score, equal error rate (EER), or false acceptance rate (FAR). To access the utility loss there is a larger variety of metrics, usually to either quantify the naturalness of the de-identified gait or to perform another kind of recognition, such as activity. One specific evaluation method we observed was by Matovu et al.~\cite{matovu_jekyll_2018} in which the authors use the biometric menagerie to observe the de-identification influence on different types of users in biometric authentication systems.

%\begin{center}
%\begin{table}
%\begin{tabular}{ c c }
%{\cont} & ~\cite{agrawal_person_2011} ~\cite{ivasic-kos_person_2014}\\
%{\discrete} & ~\cite{hoang_gait_2015} \\
%{\cont}, {\discrete} &  ~\cite{hirose_anonymization_2019} \\
%{\feature} & ~\cite{jourdan_toward_2018}\\
%{\noise} & ~\cite{tieu_approach_2017}, ~\cite{tieu_spatio-temporal_2019}, ~\cite{tieu_rgb_2019}, ~\cite{matovu_jekyll_2018}\\
%\end{tabular}
%\end{table}
%\end{center}

%\begin{center}
%\begin{table}
%\begin{tabular}{ c c c c }
%Name & Published & Participants & Source\\
%BEHAVE & 2010 & 125 & ~\cite{blunsden2010behave} \\
%OU-ISIR & 2012 & 200 & ~\cite{Makihara_CVATN2012} \\
%i3DPost & 2009 & 8 & ~\cite{gkalelis2009i3dpost} \\
%CASIA-B & 2005 & 124 & ~\cite{szhengICIP2011}\\
%\end{tabular}
%\caption{}
%\end{table}
%\end{center}

%% file: sections/hand_motions.tex
%% -*-latex-*-

\del{Hand motions are the wide variety of movements humans can perform using their hands. As they are such a universal part of human interaction with their environment there exist multiple approaches for using hand motions as behavioral biometric factors: handwriting, keystrokes, mouse movements, and hand gestures. They differ by how the hand motions are recorded and which task the person performs.
Handwriting is a hand motion in which the person performing it uses a pen to write text.  Due to the uniqueness of people's handwriting, it has long been established that humans can be identified by it. Signatures are the written name of a person which are intended for identification purposes, for example on legal documents. Handwriting can be captured offline in which the produced text is captured via a picture or online in which the movement of the pen is captured during the writing process. For this survey, we only consider the uniqueness of one writing style and not the linguistic style (Stylometry) of the written text.
In modern life, handwriting has been largely replaced by typing on a keyboard. Besides writing texts, keyboards are also used as a general input modality for computer systems. The keystrokes and the timings a human produces while using a keyboard are also a biometric factor. Another input modality that captures hand motions are mouse movements.
Hand gestures are the wide range of hand motions humans perform to communicate nonverbally. While a normal part of human communication, hand gestures only recently became important as an input modality for computer systems with the rise of swipe gestures on smartphones. This trend is continuing with freehand gestures for wearables such as smartwatches or augmented reality headsets.}

\new{We use the term hand motions as an umbrella for all hand motion related biometric factors, including handwriting, keystrokes, mouse movements, and hand gestures. These traits mostly differ by how they are recorded and what kind of hand motions are performed. Handwriting can be captured offline or online, depending on if only the resulting written text or a real-time capturing of the hand while writing is being used. For this survey, we only consider the uniqueness of one writing style and not the linguistic style (Stylometry) of the written text. In modern life, handwriting has been mostly replaced by typing on keyboards which also is an important biometric factor as individuals can be identified by the timings of their key presses. Besides keyboards also the usage of computer mice creates unique patterns, as their trajectories and clicks are again a biometric factor. Lastly, hand motions can be directly captured using optical or accelerometer tracking techniques.}

Hand motion recognition encompasses multiple recognition techniques for different capture modalities, here we give an overview of handwriting, mouse movements, keystrokes, and gestures. For handwriting bases hand motion recognition the input handwriting sequence is often adjusted for its baseline, scaled to a normal writing style, and segmented to meet the demands of the classifier~\cite{Plamondon_2000_Online}. Handwriting is further dependent if it was captured while the person was writing (online handwriting), for example with a digital pen, or only handwriting itself is captured after the person has finished (offline handwriting). The recognition for mouse movements relies on the trajectory, speed, single, and double clicks performed with a mouse as features. Keystroke-based hand motion recognition is based primarily on the timing differences between key up, down, and hold events. Besides individual events, the differences between two successive events or even three successive events are also used as features~\cite{zhong_2015_survey}. Hand motion recognition via gestures can be split into 2D gestures which are performed on a flat surface (e.g. on a smartphone) and 3D gestures which are performed in mid-air. Sherman et al.~\cite{sherman_2014_User} use the trajectories of each finger and first resamples them using a cubic spline interpolation to get a lower sampling rate, removing unwanted jitter. \del{They then use a mutual information metric to classify the gesture.~\cite{Sae-bae_2014_Multitouch} In the first step they label each finger. Then the distance between every two fingers and each finger position and its following position is calculated.} To calculate the distance between two gestures dynamic time warping is employed with various distance metrics.\del{~\cite{tian_kinwrite_nodate} The 3D gesture recognition works similar to the 2D one as first the fingertips of each finger are found and then after scaling and smoothing multiple features based on the fingertips are selected. The classification is again performed with dynamic time warping.}

\subsubsection{Utility}
The utility range for hand motions is a large and diverse field. For handwriting the resulting text must be readable either by humans or computers, the particular handwriting style is usually not important. This is different for signatures, as their main purpose is to facilitate the identification and verification of the signers identity, hence their particular style is important, while the readability of the name is less important. Since the other hand motions mostly serve as input modalities for computer systems their utility as input modality~\cite{Zhang_1998} must be kept precise and timely to keep their utility. For hand gestures~\cite{saunders_anonysign_2021}, there is additionally its utility for non-verbal communication.

\subsubsection{Threat Space}
\del{
Handwriting used to be essential to human communication but with the rise of computers, it has become less important and was mostly replaced by digital communication. Due to its decline as a communication medium, it has become difficult to get handwriting samples of a particular person. Since the sensitive nature of signatures as a biometric factor is commonly known humans are usually cautious at leaving their signature, however, due to them widely being used in everyday life there still at risk of being collected by an adversary. %Further, institutions like banks keep databases of signatures for verification purposes.
Most hand motion capturing today happens implicitly when humans use their hands to control computer systems. Each time we use a mouse or keyboard our hand motion is recorded and as such at risk. Most applications or websites could be used to capture both mouse movements and keystrokes. But even without direct access to the keyboard attackers could collect these biometrics via side-channels such as network latency.
Hand gestures are a rather new input modality for computer systems and only became widely popular with the rise of smartphones. Due to their exposed nature and the fact that we often perform gestures in public hand gestures are relatively easy to capture by an adversary, for example by using a camera. It is to be expected that with the rise of mixed reality and its applications hand gestures gain more importance as an input modality and therefore will be at a higher risk.}

\new{The threat space for hand motion is diverse as the usage of our hands is 
unavoidable in most everyday tasks and as we often use digital devices the recording of hand motions happens most of the time without us realizing it. As many studies have shown hand motions can be used to identify individuals by their handwriting~\cite{Plamondon_2000_Online}, keystroke dynamics~\cite{alsultan2013keystroke}, mouse movements~\cite{revett2008survey}, and gestures~\cite{yang2016free}. Besides identification our hand motions also often convey meaning such as when we write a text on a keyboard, the semantics of hand motions can be sensitive too, such as when we enter passwords or write private messages. Specific medical conditions manifest themselves in hand motions, such as hand tremors in Parkinson's patients~\cite{Jankovic368}. Further, hand motions convey information about our emotional state~\cite{stathopoulou2011emotion}.}

%with millions of letters send every year. With the rise of first typewriters and then computers less and less written communication is written by hand making it these days difficult to get hold of a handwritten text by a person. This hold true with the exception of human signatures which are still wildly used to verify ones identity. Capturing hand motions today often happens implicit when we use computer systems, with no prior consent being necessary. Whenever we use a keyboard, move a computer mouse, or swipe on our phone a hand motion is captured. The prevalence of computer systems in modern life makes it nearly impossible to not have our hand motions captured. Luckily, these captured are most often only processed locally on the device we use and only stored during its processing. While direct keystroke capture is difficult because an attacker would require physical access to the keyboard the inference of keystrokes via side-channels such as network traffic are easier to obtain. 

%Handwriting samples are not that simple to obtain since most text are today written on computers and people %do not leave their signature everywhere.

%Hand gestures are relatively easy to capture as they can be from afar and without the users participation. Via smartphones gestures. Via cameras. Via 

%\subsubsection{Privacy goals}

%The privacy goals are the same as in the general section.

\subsubsection{Anonymization Techniques}
In the following, we present the suitable methods for hand motion anonymization, with the exception of mouse movements as we did not find any suitable papers for it.

\paragraph{Random Perturbation}
Maiorana et al.~\cite{maiorana_bioconvolving_2011} propose a template protection method for online handwriting which splits a handwriting sequence into segments and then randomly mixes the segments before convoluting them. The same shuffling approach is taken by Maiti et al.~\cite{maiti_smartwatch-based_2016} to prevent keystroke inference attacks via wrist-worn accelerometers, however, they do not convolute the segments. The approach was only evaluated with 4 participants. Another study investigating the permutation of keystrokes is performed by Vassallo et al.~\cite{vassallo_privacy-preserving_2017}, in their evaluation they only investigate the utility reduction. Goubaru et al.~\cite{goubaru_consideration_2014} propose a template protection scheme for online handwriting templates. They extract the pattern ID for a user by using a common template. The pattern ID is then XORed with a secret that was encoded by an error-correcting code. The result is stored as the template. For the verification, the pattern ID is again extracted and then XORed with the template.

\paragraph{Noise Injection}
To prevent the identification in browsers via keystroke timings Monaco et al.~\cite{monaco_obfuscating_2017} investigate two noise injection strategies: delay mixing and interval mixing. Delay mixing adds random noise to the timing of a keystroke and interval mixing which draws a new arrival time for each keystroke, depending on a randomly drawn interval. Their results show a reduction of identification of about 40\%. A similar approach to delay mixing is also investigated by Migdal et al.~\cite{migdal_keystroke_2019} which also adds delays to keystroke timings. \new{Shahid et al.~\cite{shahid_evaluating_2021} propose to use the Laplace mechanism on the 2D coordinates of handwritten text to achieve local differential privacy.}

\paragraph{Coarsening}
Vassallo et al.~\cite{vassallo_privacy-preserving_2017} explore suppression of keystrokes to preserve the content of the typed text in a continuous authentication scenario.
Maiti et al.~\cite{maiti_smartwatch-based_2016} also focus on keystrokes privacy and propose two coarsening methods to prevent keystroke inference attacks via wrist-worn accelerometers. In their first approach, they simply detect if a user is typing via several features and then block the access to the accelerometer data to prevent attacks. Their second method reduces the sampling rate of the accelerometer.

\paragraph{Discrete Conversion}

For discrete conversion we found the following techniques aimed at template protection.
An online handwriting template protection scheme is proposed by Sae-Bae et al.~\cite{sae-bae_online_nodate} which decomposes signatures into histograms on which the authentication is performed. They use one-dimensional histograms to capture the distribution of single features and two-dimensional histograms to capture the dependence between two features. Migdal et al.~\cite{migdal_my_2019} propose a template protection scheme for multiple modalities, including keystrokes. Their scheme combines multiple pieces of information, such as ip addresses, with the keystroke information and then computes a biohash on it. 
Leinonen et al.~\cite{leinonen_preventing_2017} investigate the anonymization of keystroke timing data using two rounding approaches which effectively sort the timings into buckets. Their approach appears to be effective as the identification drops from close to 100\% to below 10\%.
Vassallo et al.~\cite{vassallo_privacy-preserving_2017} explore substitution of keys with a random nearby key to preserve the content of the typed text in a continuous authentication scenario.

Figueiredo et al.~\cite{figueiredo_prepose_2016} have developed a modeling language that can be used to design new gestures for applications. \del{Each gesture is validated to check if the gesture can harm the person performing it and if an existing gesture is overwritten by it.} The gestures can then be recognized on the recording hardware, eliminating the need to give the application access to the clear data. No privacy evaluation was performed. \new{For privacy friendly gesture recognition Mukojima et al.~\cite{mukojima_deep-learning-assisted_2022} designed a system which illuminates the hand with a random pixel pattern and captures the remaining light on the opposite site of the hand with a detector. From this reduce data collection the shape of the hand is reconstructed via machine learning. The authors did not evaluate the privacy protection of their approach.}

\paragraph{Continuous Conversion}
Maiorana et al.~\cite{maiorana_bioconvolving_2011} propose two continuous conversions for online handwriting templates: A baseline conversion which first splits a handwriting sequence into multiple segments based on a secret key and then convolutes the segments. And a shifting transformation that applies a shift to the initial sequence. The template matching is performed on the protected template. \new{For the anonymization of gestures which have been captured via inertia measurement unit (IMU) sensors Malekzadeh et al.~\cite{malekzadeh_privacy_2020} propose two separate auto encoders. The first auto encoder is supposed to replace sequences in the data which have been classified as sensitive with a generated neutral sequence. While the second one should minimize the mutual information between the data and the identity of the user. Their approach reduces the identification from 96\% accuracy down to 7\%.

Another auto encoder based approach is proposed by Saunder et al.~\cite{saunders_anonysign_2021} in which the sign language motions of one person are transferred onto another one. Their technique is two fold they first extract the pose of the source video and encode this to a set of pose features. Secondly they encode the style of the target appearance using an appearance distribution. The encoded pose and style are then combined to generate a new image. It was not evaluated if the persons can be identified by their hand motions only. A second approach to perform sign language anonymization was proposed by Xia et al.~\cite{xia_sign_nodate}. They use an estimation of the motion regions and then use optical flow in combination with a confidence map to encode the motions of the source and driving video. Then the anonymized video is generated via an auto encoder from the source video, optical flow and confidence map. To keep the utility of the sign language high they use a loss function which especially focus on the difference between hand and face motion of the driving and anonymized video. Again no evaluation if the persons can be identified by their hand motions was performed.}

\subsubsection{Evaluation}
Hand motion anonymization is mostly evaluated in the context of authentication and as such the false positive rate (FPR), false negative rate (FNR), and equal error rate (EER) are important metrics for evaluating the performance. But there is also the usage of recognition approaches for the evaluation for example by Monaco et al.~\cite{monaco_obfuscating_2017} which uses the accuracy of identity, age, gender, and handedness inference. A unique evaluation approach we found was used by Goubaru et al.~\cite{goubaru_consideration_2014} who used the randomness of the template bits via occurrences and autocorrelation to evaluate their approach.

%% file: sections/eye_gaze.tex
%% -*-latex-*-

Eye gaze involves two type of movements: \textbf{fixations} and \textbf{saccades}.
%\tc{what data? blinking, pupil diletion, direction of attention}\pc{specified below}
Our eyes alternate between them during visual tasks, such as reading\iflong (see Figure~\ref{fig:eyegaze})\else\fi. Fixations refer to maintained visual focus on a single stimulus, while saccades are rapid eye movements between fixations to reorient our gaze. %Besides, ...microsaccades?  
 Besides, even during fixations, our eyes are not completely still, but constantly producing involuntary micro movements (hundreds per second) known as microsaccades~\cite{abrams1989speed}.

%license: public domain: https://de.wikipedia.org/wiki/Datei:Reading_Fixations_Saccades.jpg

\iflong
\begin{figure}[!ht]
    \centering
      \includegraphics[width=8cm]{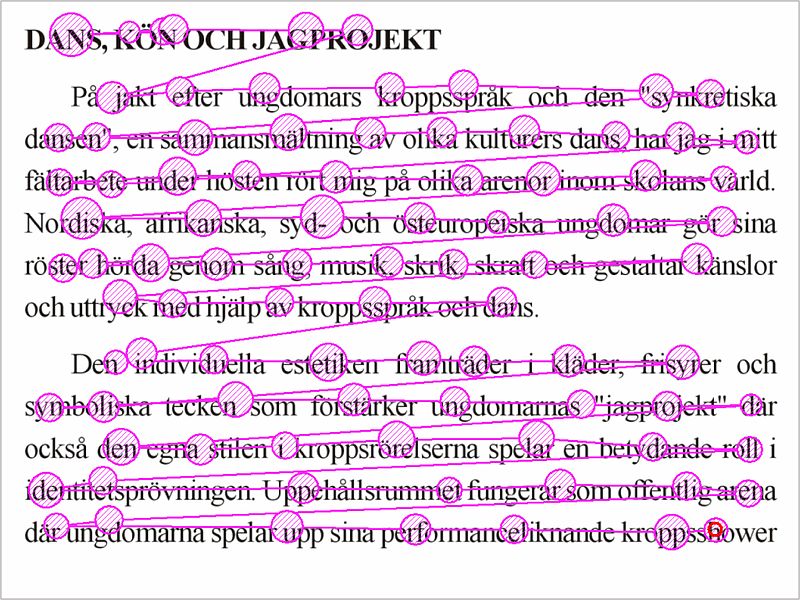}
    
\caption{Fixation and saccades while reading, from a study of speed reading made by Humanistlaboratoriet, Lund University, in 2005. Source:\url{http://en.wikipedia.org/wiki/File:Rea}.}\label{fig:eyegaze}
\end{figure}
\else\fi

% Saccades are rapid eye movements and they are considered to be the fastest rotational movement of any external part of
%our body, reaching angular velocities of up to 900 degrees per second, and usually lasting between 20 ms and 100 ms.

%This figure in Sluganovic is interesting to show uniqueness: In Figure 1, fixations can be seen as areas of large numbers
%of closely grouped points, while saccades consist of series of
%more spread recordings that depict fairly straight paths.
%\textcolor{magenta}{types of trackers: external hw, camera, glasses/VR mounted}
Eye-tracking technologies are becoming increasingly available in the consumer and research market. The most common type of tracking technology works by illuminating the eye with an array of non-visible light sources that generate a corneal reflection. These reflections are sensed and analyzed to extract eye rotation from changes in reflections. 
%Processing the relationship between this reflection and the center of the pupil, eye trackers calculate vectors that relate eye position to locations in the physical world as the eyes move~\cite{(Hansen & Ji, 2010)}.
 There is a wide range of hardware configurations for eye-tracking, including embedded cameras in computers, smartphones and virtual reality headsets, dedicated external hardware, or mobile eye-wear. These sensors allow to extract measurements not only regarding movement data related to fixations and saccades (speed, gaze angle, attention spots, scan path), but also additional features, such as pupil size variations and blink behavior. Combinations of these  features provide valuable information to implement eye-gaze driven applications.

%\textcolor{magenta}{mention Features ? and link-applications}
%https://cognitiveresearchjournal.springeropen.com/articles/10.1186/s41235-019-0159-2
%also pupil and blink behavior, will cover all types of movements and extra featuers

\subsubsection{Utility} Eye movements have been studied, analyzed and used for more than a century in different research domains. In the medical field, gaze provides useful information about our cognitive and visual processing~\cite{harezlak2018application, bahill1975main}, which can be used for diagnosing different diseases.
%\tc{Which information needs to be retained?} \pc{same as mentioned before before,different features depending on the case}
 In computer science, eye gaze is used as a form of human computer interaction to improve accessibility, user experience, and to adapt system behavior~\cite{majaranta2014eye, poole2006eye, conati2007using}. More recently, security and privacy researchers have focused on analyzing stable unique features of eye movement to build biometric authentication systems~\cite{katsini2020role}. Behavioral eye biometrics have been subject of intense investigation in the last decade, showing EERs as low as 1.8\%~\cite{eberz2018your}. Across all these different domains, the utility to be preserved would depend on the underlying application, e.g., accuracy in predicting the next eye movement, in diagnosing a mental disease, in detecting the focus of user attention, or in recognizing a user.

%For example, Castelhano
%et al. [8] examine stable individual di↵erences in characteristics
%of both saccades and fixations and provides support for
%their stable use in biometric authentication. 

%Fixations are widely analysed in human vision, gaze-based interaction, and ... human-computer interaction, **attentive user interfaces**, and eye-based user 

%Many works apply gaze-based features to build an authentication scheme, aiming to overcome the
%limitations that are introduced by the use of physiological biometrics, such as inability to revoke passwords once compromised,
%or unintentional authentications [146]. Most surveys
%considering eye gaze do not address the HCI perspective but
%focus on technical [174] and methodological [47] aspects of
%such systems (e.g., the use of bio-signals for human identification
%[41]). None of the existing surveys consider multi-factor
%gaze-based authentication, gaze-based privacy protection, or
%the use of gaze to understand users during security tasks. To
%our knowledge, this is the first survey to holistically cover the
%three major utilities of gaze in security applications
%use of gaze in security and privacy applications~\cite{katsini}

\subsubsection{Threat Space} Eye movement data is rich in information that can be exploited by malicious entities or curious service providers to uncover user sensitive attributes beyond those disclosed intentionally and required for the purpose of the service or to directly identify a person. %\textcolor{magenta}{ diseases, some attack examples.}
  Besides the biometric information carried by eye movement data, research has also documented their correlation with multiple disorders and mental conditions, such as Alzheimer’s~\cite{hutton1984eye}, schizophrenia~\cite{levy2010eye, holzman1973eye}, Parkinson~\cite{kuechenmeister1977eye} bipolar disorder~\cite{garcia2014attentional}, mild cognitive impairment~\cite{yang2013specific}  multiple sclerosis~\cite{derwenskus2005abnormal}, Autism~\cite{boraston2007application, wang2015atypical}, or psychosis~\cite{ettinger2004volumetric}, to name a few. Furthermore, pupil size is known to be an indicator of a person’s interest in a scene~\cite{hess1960pupil} and a proxy for detecting cognitive load~\cite{matthews1991pupillary, krejtz2018eye}. Other recent works demonstrated that eye data can be used to infer gender and age, or even personality traits~\cite{kroger2019does, berkovsky2019detecting}. Given the richness of eye data and the increased availability of consumer tracking devices and the advent of eye-gaze driven applications,  there is a significant and imminent privacy threat potential~\cite{adams2018ethics}.
 
 %Otherworks have shown that eye movements are closely
%linked to mental disorders, such as Alzheimer’s [Hutton et al. 1984],
%Parkinson’s [Kuechenmeister et al. 1977], or schizophrenia [Holzman
%et al. 1974]. More recent work in HCI has demonstrated the use
%of eye movement analysis for human activity recognition [Bulling
%et al. 2013; Steil and Bulling 2015] as well as to infer a user’s cognitive
%state [Bulling and Zander 2014; Faber et al. 2017] or personality
%traits [Hoppe et al. 2018]. More closely related to our work, several
%researchers have shown that gender and age can be inferred from
%eye movements, e.g. by analysing the spatial distribution of gaze
%on images like faces [Cantoni et al. 2015; Sammaknejad et al. 2017].
% more: https://www.sciencedirect.com/science/article/pii/S0895611117300435

The two main threats that endanger eye privacy are re-identification and attributes' inference. %, which lead to the  

%There is an eye movement identification and verification competition

%\subsubsection{Privacy Goals} To protect against the above specified threats, the privacy goals of anonynmization techniques to protect eye gaze data are de-identification and attribute removal.

\subsubsection{Anonymization Techniques}
%\textcolor{magenta}{--14 papers, 3 of them do differential privacy, 
%2 of the DP are very similar: add laplace and (fourier ?) noise and apply it
%to document type detection without gender inferences
%2nd paper cuts in chunks and de-correlates the feature before adding noise
%the 3rd DP paper protects heatmaps but it does not provide an utility metric
%just a visualization}

\del{We found three proposals to protect the privacy of eye movement data~\cite{liu2019differential, bozkir2020differential, Steil19ETRA}, all of them are guided by differential privacy (DP). The general idea of differentially private algorithms is to add a certain amount of randomly generated noise to the original signal, so that it is difficult to say whether or not an individual contributed their data.}

\new{We found multiple recent proposals to protect the privacy of eye movement data, with many of them using noise injection to achieve differential privacy (DP).}

%the data from any two individuals is essentially indistinguishable~\cite{dwork2014algorithmic}.\jp{More precisely, ``so that it is difficult to say whether or not an individual contributed their data''.}

\new{
\paragraph{Random Perturbation}
David-John et al.~\cite{david-john_for_2022} adapt the task-based marginal model for eye gaze, in which for each feature vector dimension a distribution of the values is build to then random sample new synthetic data from these distributions. The identification accuracy of the generated synthetic data is close to chance level.
}

\paragraph{Noise Injection} Steil et al.~\cite{Steil19ETRA} propose a DP-based technique to protect eye movement data collected while users read different types of documents (comic, newspaper, textbook) in a VR setting. The utility goal is to accurately predict the type of document to provide enhanced features in the reader application. Additionally, the privacy goals are to avoid gender inferences from eye movement data and to protect against re-identification when the attacker has prior knowledge of a data set including the target user eye data and identity. To achieve these goals, the exponential mechanism~\cite{dwork2014algorithmic} is applied to a database of users' eye features by a trusted curator prior to its release. This sanitised database can be then used for training classifiers to provide the enhanced reader functionality. The experiments testing at various noise level show that utility with regard to document classification can be partly preserved ($\sim$55-70\%) while reducing gender accuracy inference to the level of random guesses ($\sim$50\%).

%---noise proportional to eye feature variations during the reading task 
%noise is reading task specific
%Threats are "mitigated"

Based on Steil et al.'s data set, Bozkir et al.~\cite{bozkir_differential_2021} evaluate two types of DP-based perturbations, the standard Laplacian perturbation algorithm (LPA)~\cite{dwork2006calibrating} and the Fourier perturbation algorithm (FPA)~\cite{rastogi2010differentially}. They also propose a modification of the FPA algorithm that splits eye data in chunks before adding noise, in order to reduce temporal correlations, which is a source of reduced utility as more noise is required to protect privacy. With this modification, they obtain document type classification results similar to those used by Steil et al.~\cite{Steil19ETRA} for the case of 50\% gender classification, while adding more noise to the data (better privacy guarantee).

%--> similar results at epsilon 4.8 than Steil at epsilon 15. Unclear the epsilon interpretation

Liu et al.~\cite{liu2019differential} present a DP-based solution to anonymize eye tracking data aggregated as a heatmap. A heatmap, or attentional landscape, is a popular method for visualizing eye movement data that represents aggregate fixations~\cite{duchowski2017eye}.
 %[Duchowski 2018].Pomplun et al. [1996], Wooding [2002],Duchowskiet al. 2012].
  This means that the intensity of every pixel is adjusted relative to the number of fixations over that region. The privacy goal in this case is to protect individual gaze maps while preserving the utility of the aggregated heatmap. Their experiments with random selection and additive noise (Gaussian, Laplacian) show that Gaussian noise is the best option to obtain good privacy guarantees for the individuals' gaze maps without visually distorting the hotspots in the aggregated heatmap, i.e., keeping a certain utility.

\new{David-John et al.~\cite{david-john_privacy-preserving_2021} worked on protecting eye tracking data recorded in VR/AR headsets. They propose two different interface models how data can be shared with a third party and propose three anonymization techniques, Gaussian noise injection, temporal down sampling, and spatial down sampling for on of the interface models. The noise injection approach was found to be the most effective as it reduced the identification rate of the subjects the most with high variance values for the Gaussian distribution.
}

\new{Hu et al.~\cite{hu_otus_2022} proposed a local differential private mechanism for generating synthetic eye movement trajectories called Otus. Their technique first separates the field of view into tiles and then constructs a graph which encodes the gaze duration of each tile and the transition probability between the tiles. The graph is then perturbed using the Laplacian mechanism before it is send to the server. The server then averages all users graphs and uses random walks on the graph to generate new eye movement trajectories.
}

\new{Li et al.~\cite{Li_2021_Kaleido_USENIX} proposed Kal$\epsilon$ido a plugin system which can be used to anonymize eye gaze trajectories with differential privacy guarantees. The authors extend geo-indistinguishability~\cite{andres2013geo} and w-event privacy~\cite{kellaris2014differentially} to take into account the area of interest with radius r a user is looking at. The intuition of their guarantee is that all gaze positions within the area are indistinguishable. They note that they only protect against spatial information and not temporal information. Further, they define an adaptive algorithm to allocate the privacy budget of a user depending on the total privacy budget of each time window. Their results show a reduction of the identification of users to near chance level, however the utility of the data is also close to chance level.
}

\new{
\paragraph{Coarsening}
The temporal and spatial down sampling proposed techniques by David-John et al.~\cite{david-john_privacy-preserving_2021} are both coarsening base techniques. For the temporal down sampling only a very small reduction in the identification accuracy can be recorded while the spatial down sampling has a bigger effect but must be scaled very high to do so.}

\new{
\paragraph{Continuous Conversion}
David-John et al.~\cite{david-john_for_2022} applied k-anonymity to eye movements by grouping the trajectories of users and then averaging them. They were able to show that even with small numbers of k the identification accuracy drops significantly. Due to them processing the feature vectors of each task separately their reported high utility is questionable.
}

\new{
Fuhl et al.~\cite{fuhl2021reinforcement} perform eye gaze anonymization by using an auto encoder in combination with reinforcement learning. The auto encoder is trained on the eye gaze trajectories to learn a latent representation of the data. Then a manipulation agent modifies the latent vector of the trajectories to prevent for example gender classification. After the decoding of the latent vector a classifier tests how good the manipulation was and its result is used as the loss for the training of the manipulation agent.
}

\subsubsection{Evaluation} The proposals by Steil et al.~\cite{Steil19ETRA} and Bozkir et al.~\cite{bozkir2020differential}, measure the quality of their anonymization techniques for attribute inference protection using the classification accuracy metric for the main task and the attribute inference task. For the re-identification protection case, it is assumed that the attacker has previous knowledge of a database
%\sh{clear database?}
 of users' eye data and their identities. To simulate this knowledge, they train the classifiers on the clean data and test them on the anonymized data
 %\sh{Maybe "anonymized data"}
 , using also the accuracy metric to report privacy protection. Besides, these works also report the so called privacy loss parameter (or $\epsilon$) from DP theory, which quantifies the maximum difference between the data points of two individuals in the data set. Furthermore, Bozkir et al. use the inverse of the normalized mean square error (NMSE) between the actual eye feature values and the perturbed ones as a utility metric. However, the interpretation and implications of these privacy loss and utility metrics are not developed. 
%TODO: As the chunk size decreases, the chunk level sensitivity decreases as well as the computational complexity.
%However, the parameter ǫ′ that is calculated according to the correlation level
%becomes larger with smaller chunk sizes due to the fact that correlations between
%neighboring data elements are larger in an eye movement datasetdataset. Therefore, it
%is important to obtain a good trade-off between computational complexity and
%correlations to determine the optimal chunk size.
%NMSE

Liu et al.~\cite{liu2019differential} analyzed the privacy-utility trade-off of anonymized heatmaps using the correlation coefficient (CC) and mean square error (MSE) of noisy heatmaps under different privacy levels (different values of $\epsilon$). The CC and MSE give an idea of the similarity between the original and the anonymized heatmaps and the $\epsilon$ provides information about the privacy guarantee (the smaller, the better privacy). These metrics are accompanied by the visual representation of the noisy heatmap, in order to aid the relevant stakeholders in deciding what level of noise is acceptable for a given application. 

Regarding datasets, Steil et al.~\cite{Steil19ETRA} collect  data from 20 participants (10 male, 10 female, aged 21-45) while reading documents using a VR headset. Each recording is divided into three sessions (reading a comic, newspaper, or textbook), lasting 30 minutes in total. They extract 52 eye movement features related to fixations, saccades, blinks, and pupil diameter. The dataset has been publicly released~\footnote{https://www.mpi-inf.mpg.de/departments/computer-vision-and-machine-learning/research/visual-privacy/privacy-aware-eye-tracking-using-differential-privacy} by the authors and Bozkir et al.~\cite{bozkir2020differential} use it as the basis to evaluate their proposal. In the heatmaps anonymization study, Liu et al. use a synthetic simulated dataset to illustrate their privacy analysis. Besides the technical privacy analysis, Steil et al.~\cite{Steil19ETRA} is one of the few works considering user privacy concerns regarding behavioral data collection. They conduct a large scale user survey (with N=164 participants) to explore with whom, for which services, and to what extent users are willing to share their gaze data. Their report shows that people are uncomfortable with inferences (gender, race, sexual orientation) and would object to share their data if these attributes can be leaked. The results also show that people generally agree to share their eye tracking data if a governmental health agency or for research purposes, but would object to do so if the data owners are companies. These insights are a first step towards understanding user privacy awareness and privacy needs, but more work is required in this field to guide the design of user-centered privacy protective techniques for behavioral data.

%3 does not provide an utility metric just a visualization, simulated heatmaps
%It would be interesting and important to define and analyze differential privacy for temporal data
%are needed especially on the data sharing side of eye tracking. I

%% file: sections/heartbeat.tex
An electrocardiogram (ECG) is a graph of voltage over time that captures the electrical activities of cardiac muscle depolarization followed by repolarization during each heartbeat. \iflong Shown in Figure~\ref{fig:heartbeat}, the\else  The \fi  ECG graph of a normal beat is composed of a sequence of waves: a P-wave reflecting the atrial depolarization process, a QRS complex representing the ventricular depolarization process, and a T-wave denoting the ventricular repolarization. Other portions of the ECG signal encompass the PR, ST, and QT intervals~\cite{ZhengSC20}.

\iflong
\begin{figure}[!tb]
\centering
\includegraphics[width=5cm]{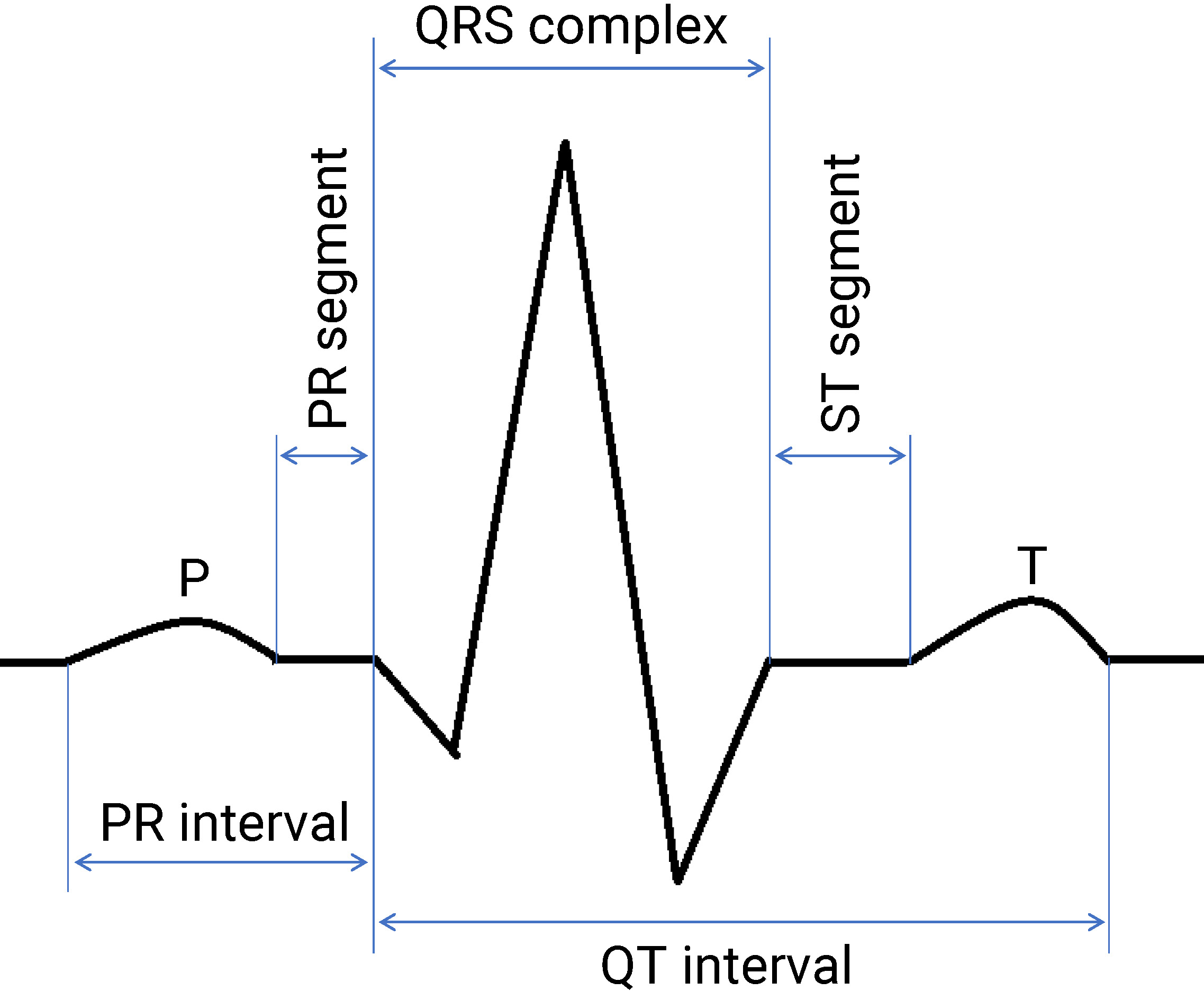}
\caption{Waveform of an ECG signal with normal cardiac cycle. Source: \url{https://www.nottingham.ac.uk/nursing/practice/resources/cardiology/function/normal_duration.php.}}
\label{fig:heartbeat}
\end{figure}
\else\fi

\del{As per the current screening and diagnostic practices, cardiologists review ECG data, find the right diagnosis and implement subsequent treatment plans such as a medication regime or the removal of a radiofrequency catheter. Nonetheless, the demand for highly accurate, automated heart-condition diagnoses has increased significantly, in part due to the new public health regulations of implementing more extensive screening processes as well as the adoption of ECG-enabled wearable devices.}

\del{It is well known that certain types of cardiovascular conditions, such as atrial fibrillation, have a wide and severe impact on public health, quality of life, and medical expenditures. The long-term ECG monitoring is a vital, non-invasive tool for detecting such conditions. For evident computational and intellectual property reasons, however, the analysis of such data is usually not conducted at the wearable device but at automated, machine-learning based systems typically hosted in hospitals or external service providers. This necessarily implies the transmission of ECG data from patients to non-fully trusted entities, which inevitably poses evident privacy risks.}

\del{However, the disclosure of sensitive data not only represents a threat to patients’ privacy: it may also prompt a serious security risk to any biometric-authentication system that relies on those data. The advantage of ECG-data-based systems over other biometrics systems (like fingerprint, face or iris), though, is the intrinsic nature of ECGs and also their inherent indication of life, which make them very difficult to forge or steal~\cite{HB28}. Compared to fingerprint and facial recognition systems, where extra sensors ---other than those required for medical monitoring purpose--- are needed, ECGs are a more suitable choice in practical applications and have been shown to be extremely accurate in identification tasks~\cite{SinghBIO08}.}

Like other biometric systems applied to identification tasks, ECGs are typically converted into abstract, compressed representations, typically referred to as biometric templates\del{\footnote{As already mentioned in previous subsections, the functioning principle of biometric templates is that an original signal can be recovered from its template.}}, before the task is conducted\del{\footnote{Bear in mind that ECG signals are generally collected over long periods of time and at high resolutions. This leads to large volumes of data being collected. For example, for a sampling rate of 500 Hz and a data resolution of 8 bits per sample, a 24-h record amounts to about 43.2 Mbytes per channel.}}. Biometric-template methods can be classified depending on the exploited features of the ECG data. The most popular ones are fiducial-based, non-ﬁducial-based and hybrid methods~\cite{OdinakaTIFS12}. On the one hand, fiducial-based techniques utilize characteristic points on the ECG signal to extract temporal, amplitude, envelope, slope and area features. Characteristic points are the locations that correspond to the peaks and boundaries of the P, QRS and T-waves of the ECG signal. On the other hand, the non-ﬁducial-based methods do not rely on the ECG characteristic points, and examples include autocorrelation coefﬁcients, Fourier and wavelet transforms. Hybrid methods combine both ﬁducial-based and non-ﬁducial-based features.

\del{Biometric templates are therefore an attempt to reduce data storage in identification services. In other type of services, ECG signals are expectedly compressed to allow efficient transmission and storage as well. As we shall elaborate later, techniques aimed to protect the transmission of ECG data will be classified depending on whether they are applied before or after compression.}

\subsubsection{Utility} 
ECG data find application in healthcare and biometrics systems, the latter being intended for identification and authentication\new{~\cite{Uwaechia_2021}} \del{, as discussed in the preliminaries of this section}.
In healthcare, ECGs are utilized for  \new{diagnosis of heart diseases~\cite{LIU2021107187}}\del{remote diagnosis and in-home health monitoring}. Typically, there is a stand-alone service or a complete e-health system where the service provider, in addition to offering a repository of personal medical data, may allow to remotely process such data. In any case, the aim is to provide real-time feedback to patients and hospitals, either as a warning of impending medical emergency or as a monitoring aid during physical exercises.

\del{Although it is well known that ECG data may help diagnose a patient’s physiological or pathological condition, other probably lesser-known inferences include cocaine use~\cite{HossainIPSN14} and stress~\cite{PlarreIPSN11}, which may be sensitive to the patient and obviously should be kept private. The fact that the very same time series data allows drawing both desirable inferences (i.e., for healthcare) and sensitive inferences (that need to be protected) poses a dilemma of great practical relevance.}

\subsubsection{Threat Space} Regardless of the application (i.e., identification, authentication or healthcare), ECGs are health data and, as such, are considered sensitive by data-protection regulations and need to be protected. Consider the case, for example, of a user who might see their insurance premium increased or suffer discrimination during a job application due to a medical condition inferred from their ECGs.

\new{Although it is well known that ECG data may help diagnose a patient’s physiological or pathological condition, other probably lesser-known inferences include cocaine use~\cite{HossainIPSN14} and stress~\cite{PlarreIPSN11}, which may be sensitive to the patient and obviously should be kept private. The fact that the very same time series data allows drawing both desirable inferences (i.e., for healthcare) and sensitive inferences (that need to be protected) poses a dilemma of great practical relevance.}

\del{The general scenario where ECG data are collected\iflong, processed and stored is shown in Figure~\ref{fig:heartbeat:threat}\else\fi. The scenario is composed of three entities: a patient (in the case of healthcare applications) or user (in the case of identification and authentication applications); a wearable or internal\footnote{In the sense of within a patient or user’s premises.} device collecting patient’s ECG data; and an external entity that receives the data collected by the internal device, and processes and stores such information as a biometric template or in raw or compressed format, so as to provide a service.}

\del{
\iflong
\begin{figure}[!tb]
\centering
\includegraphics[width=6cm]{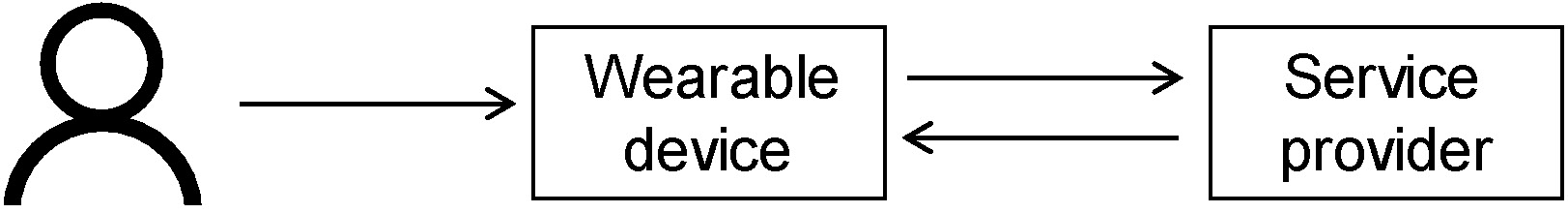}
\caption{Key entities in a scenario where ECG data are collected, processed and stored.}
\label{fig:heartbeat:threat}
\end{figure}
\else\fi
}

\del{Although the internal device is typically assumed to be trusted, both this and the external entity may be either trusted, partially trusted or fully untrusted. In the latter two cases, at the external entity the access of ECG data (including biometric templates) by unauthorized personnel poses an evident privacy threat and therefore should be prevented. As we shall elaborate in the coming subsections, privacy-protecting techniques will need to be put in place to allow only authorized personnel (e.g., medical personnel or cardiologists) to have access to ECG data or be able to reconstruct them from a biometric template.}

%In addition to the three entities described above, passive and/or active adversaries are also part of the threat space of ECG data. On the one hand, the former adversaries may eavesdrop the communications between the patient and the internal device, and between the latter and the external entity, trying to capture and decode the transmitted data with off-the-shelf or custom-built radio equipment. On the other hand, active adversaries may send radio commands, or replay recorded control commands, to a wearable device, aiming at triggering data transmission from the internal device or modifying its settings~\cite{ZhengWPMC14}. \sh{For me this is more about the security of communication then the privacy threat, due to our publishing scenario we anyway assume that the attacker get control of the data.} \jp{Removed.}

\del{Another aspect to consider within the spectrum of potential threats is the algorithm itself used by the service provider (e.g., a company or a hospital) to process ECG data. We have already mentioned that patients or users of the service could see their privacy compromised if their personal signal information or their biometric template was disclosed to a non-fully trusted-third party (not necessarily a hospital or doctor). However, protection is not only required by patients or users, but also by the service provider itself, which may not be willing to provide the end-user with its proprietary protocols because of fear of disclosing valuable intellectual property to third parties or compromising the basis for its service~\cite{LazzerettiMS12}.}

\del{To conclude, a service provider might want to make the inferences model available to any health professional, e.g., through controlled queries, and/or like to publish anonymized ECG data as a means to crowdsourcing algorithmic development\footnote{The Netflix Prize~\cite{Bennett07netflix} is probably the best-known example of collaborative-problem solving in the computer-science community.}. In both cases, the threats space would include the learned model and the released or published data.}
%, which in principle could be accessed by any adversary.

%\subsubsection{Privacy Goals.} Exactly as with other behavioral traits, the privacy goals of anonymization techniques for ECG are the protection against re-identification and against inferences beyond the information necessary to provide the service.
\subsubsection{Anonymization Techniques}
Next we survey the most relevant privacy-protection techniques for ECG data.

\del{\paragraph{Random Perturbation} As mentioned in the preliminaries, large volumes of data are collected in ECG-monitoring applications, and compression is very often needed for their transmission and storage. In this sense, Liu et al.~\cite{HB31} propose combining compression and encryption to provide privacy and confidentiality. Their proposal, however, differs from the typical compression-then-encryption approach, which may be problematic when untrusted network providers may conduct the compression task but do not have access to the private keys. The encryption-then-compression technique proposed by Liu et al. is composed of two steps. First, the ECG data, which are stored in a matrix, are multiplied by an orthogonal, randomly-generated key matrix. Then, singular-value decomposition (SVD) ---a popular dimensionality-reduction technique--- is applied to the encrypted data to provide compression.}

\del{
Another approach based on compressive sensing (CS)~\cite{CompressedSensing} is proposed by Djelouat et al in~\cite{HB9}. CS is a signal processing technique that combines both sampling and compression through random projections.
Building on this technique, the authors propose compressing the ECG signal by sampling it at the time of sensing. This reduces the need to even store the sensitive ECG data at the wearable device, thereby providing protection against that entity. The theoretical properties of this compression technique ensure that, under certain assumptions on the random projection, a good reconstruction of the original ECG signal can be obtained at the provider side.}

\sh{I dont see how these compressed sensing schemes realy fit our model. Most schemes look more like security schemes as they protect the transmission}

\paragraph{Feature Removal} Kalai et al.~\cite{HB12} present a \new{template protection scheme for ECG data.}\del{solution to secure the transmission of the ECG template between a wearable device and a service provider.} In a first phase, the authors propose computing the discrete cosine transform (DCT) of the ECG signal’s autocorrelation coefficients, and then removing those DCT coefficients with the lowest energy. The remaining DCT coefficients constitute the biometric template. In a second phase, two keys are obtained from the template. One is transmitted to the target application the user wishes to authenticate. The other functions as a private key, which is derived from the complete DCT already stored in the server. 
A similar approach is presented by Zaghouani et al.~\cite{HB13} that uses a quantization step once the DCT-template is obtained. This latter approach is evaluated on the PTB dataset but no experimental comparison is conducted between the two proposed solutions.

Another similar proposal is made by Mahmoud et al.~\cite{HB17}, which decomposes the ECG signal into its wavelet transform, eliminates the low-frequency coefﬁcients and reconstructs the ECG signal for release. At the provider side, only authorized personnel with access to a secret key (derived from the wavelet-transform template) is able to reconstruct the original ECG from the released, protected signal. To which extent these released data may safeguard patients’ privacy is evaluated through the percentage root mean square difference (PRD), a simple and widely used distortion measure in ECG signal processing applications~\cite{ECGdistortion} that quantifies the difference between the original ECG and its protected version.

\del{
Utilizing the same transform, \cite{HB25}~proposes that, after the decomposition, the essential parts of the coefficients (which consists in the P, QRS and T signatures of the ECG) are treated differently, as follows. The non-essential parts of the signal are uploaded to a public repository in the clear, whereas the essential parts are encrypted and distributed among the healthcare experts in charge of analyzing patients’ ECG data. In this process, the encrypted essential coefficients act as a key to reconstruct the original ECG, which can only accessed by authorized personnel.}

\sh{This last one does not fit our model as it uses encryption to protect who can access the data, and does not transform the data into an anonymized form}

\paragraph{Continuous Conversion}
\new{
Bennis et al.~\cite{bennis_application_2021} proposed a simple k-anonymity scheme for ECG data. In their first step they transform the signal into the frequency domain. Next they pick the k closest neighbours of the signal and then aggregate those into a new signal before transforming it back into the time domain.

Piacentino et al.~\cite{piacentino2020generating} used a GAN to generate synthetic ECG data by first normalizing the data and then arranging it into a matrix. For the arranging of the data multiple proposals are made sorting the data values by their type. No evaluation of the privacy of the synthetic data was performed.
}

\paragraph{Random Perturbation + Noise Injection}
Although encryption based on the idea of CS can achieve a computational notion of secrecy through the random projection step, it has been shown this technique is vulnerable from an information-theoretic perspective~\cite{CSIP}. To address this problem, Chou et al.~\cite{HB7} propose using principal component analysis and SVD on a CS scheme, where the ECG data is encrypted at the wearable sensor by adding signal-dependent noise. They measure privacy as the mutual information between the original ECG signal and its encrypted version, and show that high classification accuracy can be achieved while providing privacy beyond computational secrecy.

\sh{Is this encryption or anonymization?}

\paragraph{Discrete Conversion + Noise Injection} Unlike the works surveyed previously, the goal of Zare-Mirakabad et al.~\cite{HB32} is to publish suitable representations of ECG data with certain privacy guarantees. To do this, Zare-Mirakabad et al. propose converting ECG time series into symbolic representations over time. They use the popular Symbolic Aggregate approXimation (SAX) to replace continuous numerical values with strings of symbols\iflong (see Figure~\ref{fig:sax})\else\fi. With this new symbol representation, the proposed anonymization technique first builds an n-gram model from the complete time-series string, and then ensures that each n-gram has a minimum frequency of occurrence, similar to the $k$-anonymity criterion. To ensure this version of $k$-anonymity is satisfied over the string of symbols, the authors contemplate adding fake n-grams to the original string. Experimental results on the Eamonn Discord Dataset show that (a measure of) information loss is hardly affected for values of $k$ up to 20.

\paragraph{Continuous Conversion + Random Perturbation} Chen et al.~\cite{HB5} and subsequent work by Wu et al.~\cite{HB28}, address the problem of making ECG-based biometric templates revocable, exactly as keys or passwords, a property they consider indispensable in order for ECGs to be used in practice. To enable template revocability, the common practice is to associate distinct templates with the same biometrics by perturbing them in a different manner. To protect user privacy, however, this process needs to ensure the recovery of the original biometric from its template is either infeasible or computationally hard.

Essentially, cancelable templates are obtained as random projections of a user’s ECG data block. Unlike common approaches, however, Wu et al. put no restrictions on the generator matrix. Accordingly, the idea is that each realization of this matrix allows cancelling their corresponding templates. Reidentification is then conducted with the multiple-signal classification algorithm~\cite{MUSIC}, reporting rates of over 95\% in the Physikalisch Technische Bundesanstalt Database.

A distinct approach by Hong et al.~\cite{HB10}, proposes a template-free identification system to prevent any privacy issue from compromised or stolen templates. The system converts ECG-data into images through various spatial and temporal correlations methods and uses deep-learning techniques to train a classifier. The authors conduct experiments on the Pysikalisch-Technische Bundesanstalt database and report identification rates of over 90\% with sampling rates of 1\,000 Hz.

\paragraph{Continuous Conversion + Noise Injection} Sufi et al.~\cite{HB24} propose building templates of the waves P, QRS and T through cross-correlations of the ECG signal. Each of those templates are then obfuscated in a concatenated fashion with additive noise generated synthetically, so that the obfuscation of a wave serves as input to obfuscate the next wave. The upshot are noisy forms of the three waves and noisy templates thereof. All this information constitutes the key available to authorized personnel, who will be able to reconstruct the original ECG from the noisy version (which is shared or made publicly available by the patient or user themselves). Unauthorized personnel, per contra, will only have access to the noisy ECG signal, which, according to the authors, may prevent identity and attribute disclosure.

\del{
Chen et al.~\cite{HB6} tackle the problem of federated learning, where the goal is to train a machine-learning classifier with ECG data distributed over a set of entities (e.g., health institutions). The authors assume the central server coordinating the learning process and updating the global model is untrusted and resorts to block-chain technology to address this issue. Differential privacy is the privacy model used to guarantee the privacy of entities’ patients. Specifically, the authors rely on the common approach of adding noise to the local gradients, and address the asynchronous problem that arises when local gradients are missing or delayed in each iteration, by adopting the solution proposed in~\cite{ZhengAS16}. Experimental results on the MIT-BIH ECG Arrhythmia Database~\cite{MIT-DB} show the classification performance over ten types of cardiac arrhythmia is around 20\% in test error for $\varepsilon=0.1$ and 600 iterations.
}
\sh{Removed because its federated learning}

Huang et al.~\cite{HB11} propose an authentication system that protects the privacy of ECG templates in a database with differential privacy. The authors assume the interactive setting of this privacy notion, where an analyst queries the database to obtain ECG data. Specifically, the analyst is supposed to ask for the coefficients of a Legendre polynomial, that the anonymization system utilizes to fit and compress the ECG signal. Laplace noise is calibrated to the sensitivity of those coefficients and added to them, and the noisy response is returned to the analyst. The $\varepsilon$ parameter of DP therefore regulates the trade-off between user privacy and authentication accuracy, the latter aspect depending on two sources of error: the polynomial fitting approximation and the injected noise. The authors evaluate the system in the MIT-BIH ECG and MIT-BIH Noise Strees databases, reporting decent authentication accuracy. However, they appear to misunderstand how the sensitivity of the coefficients is computed and therefore their results seem to have been obtained incorrectly.

 Saleheen et al.~\cite{HB21} investigates if sensitive inferences from segments of time series data can be drawn by a dynamic Bayesian network adversary. The adversary is assumed to estimate a range of behavioral states about the user, including, for example, whether or not they are in a conversation, running, smoking and stress, at the time the data is gathered. When the adversary is likely to infer sensitive aspects of a user, the corresponding segments of data are substituted for most-plausible, non-sensitive data. To estimate the privacy provided by these substitutions of data, the authors propose a variation of the differential-privacy notion that bounds the information leaked resulting from the substitutions. In other words, the proposed metric ensures that the information leaked about a sensitive inference from a substituted segment is always bounded. Utility loss is, on the other hand, computed as the absolute difference between the probability of inference about each non-sensitive behavioral state from actual data, and the same probability from released data. Although experimental results show relatively small values of utility loss for $\varepsilon\in [0.05, 0.65]$, the proposed solution has two main limitations: first, protection is provided only for dynamic Bayesian network adversaries; and secondly, it assumes all time-series data are available beforehand, which precludes its application in real-time scenarios.

Delaney et al.~\cite{HB8} investigate the ability of generative adversarial networks (GANs) to produce realistic medical time series data. Typically, the access to medical data is highly restricted due to its sensitive nature, which prevents communities from using this data for research or clinical training. The aim of this work is to generate synthetic ECG signals representative of normal ECG waveforms without concerns over privacy. On the one hand, the authors measure utility as maximum mean discrepancy (MMD) and dynamic time warping (DTW), two common approaches to estimate the dissimilarity between two probability distributions and two time series, respectively. On the other, user privacy is evaluated as the accuracy of a membership inference attack who strives to ascertain whether or not a user’s data was used for training. Experimental results on MIT-BIH Arrythmia Database~\cite{MIT-DB} show that MMD favours GANs that generate a diverse range of outputs, while DTW is more robust against training instability. Although the authors report low accuracy results for such inference attacks, it is unclear if their solution would protect against more recent, sophisticated~\cite{MIA_CISPA} versions of those attacks.

\iflong
\begin{figure}[!tb]
\centering
\includegraphics[width=6cm]{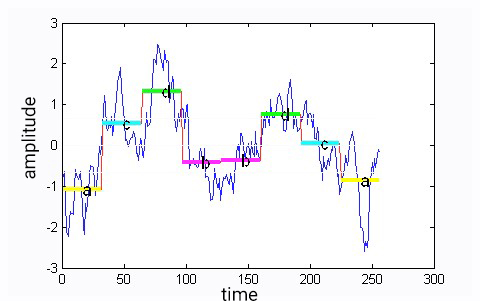}
\caption{A time series is converted into the string ``acdbbdca''. Source: \url{https://cs.gmu.edu/~jessica/sax.htm}.}
\label{fig:sax}
\end{figure}
\else\fi

\subsubsection{Evaluation} The reviewed techniques measure how service functionality is degraded due to anonymization with common machine learning metrics like precision, recall and accuracy, and less frequently with the DTW and PRD quantities, which assess the similarity between original and protected time series.
As for privacy, the level of protection is assessed through a variety notions and measures, including the accuracy of a membership inference attack, the $\varepsilon$ parameter of differential privacy, the mutual information between the original ECG signal and its encrypted version, the probability of correct inferences on sensitive attributes with and without protection, and through a notion similar to $k$-anonymity.

%% file: sections/brain_activity.tex
Brainwaves are patterns of measurable electrical impulses emitted as a result of the interaction of billions of neurons inside the human brain. Since the first human electroencephalogram was recorded in 1924~\cite{haas2003hans}, both the hardware devices to measure brain activity and the analysis techniques to process these signals have significantly improved. %\textcolor{red}{Devices, components (alpha, beta, gamma...),  analysis techniques (EEG-BCI, ERP)}
%ERP -- wikipedia
%invasive non invasive: Karen's thesis "How to capture brain signals"
%Over time, several BCI methods usable to obtain brainwaves from the human brain have been developed.
 Current technologies to measure brainwaves can be classified as invasive and noninvasive methods. Invasive methods record signals within the cortex by directly implanting electrodes near the surface of the brain \cite{KKHM+17}. These methods are far too risky for usage under noncritical circumstances and only used in clinical applications.
 %such as authentication. %Invasive measures, e.g. electrodes implanted into the brain or electrocorticography (ECoG), can provide improved scan results but are comparatively infrequent due to the high effort and risk of operations \cite{Bonaci}.
 Instead, non-invasive methods are most frequently used and applicable to many areas other than the medical realm, such as brain-controlled interfaces. The most portable and commonly used of these techniques is electroencephalograpy (EEG), which records electrical activity through sensors placed on the scalp surface.
 
 An EEG signal is a combination of different brainwaves occurring at different frequencies. Every type of wave carries different kinds of information, which can be used to gain insights about the current state of the brain \cite{AE13}. Researchers have tried to identify certain mental states associated to each brainwave.\iflong Table \ref{tab:waves} presents a summary of the most important wave types, their respecttive frequencies, their originating location in the brain, and their associated mental state.\else\fi
%\tc{What kind of data? time-series, collected passive, upon trigger?}
 
 Brain-computer interface (BCI) technologies mostly work on continuous EEG data recordings, i.e., time series data. But there are also many applications based on  the extraction of time-locked brain variations that appear in reaction to external stimuli. These variations, called event related potentials (ERPs), are widely used to detect neurological diseases. In both cases, either using ERPs or a longer EEG series, features are computed for the brainwave data-driven application built on top. These features can belong to the time and/or frequency domain, and to one or multiple channels. Examples of commonly used features include Autoregressive coefficients, Fourier and Wavelet transforms. 

%Then various computational features are extracted. More often, the channel/feature selection and dimensionality reduction methods are adopted to reduce the dimensionality and computational complexity of the EEG data, which are typically large-dimensiona EEG signals because how well the extracted features represent the EEG signals influences the performance of the recognition system. The adopted features can be classified into groups of domains
%(i.e., time domain, frequency domain, and time-frequency domain) or channels (i.e., single-channel
%and two-channel). Here, the widely used features in time domain, frequency domain, and timefrequency
%domain are described.
%Power SpectralDensity (PSD), Wavelet Packet Decomposition (WPD), 

\iflong

\input{sections/table_waves}

\else\fi

\subsubsection{Utility} The utility that should be preserved when processing brainwave data is highly dependent on the application. For clinical applications, for example, the raw information could be needed for a proper diagnosis or a safe brain controlled prosthesis. In these cases, regulations like the HIPAA Privacy Rule~\cite{hipaa} are usually in place to protect personal identifiable information. When moving to other less regulated fields of application, the need for full raw EEG data is not necessarily justified. The most prominent EEG applications include user authentication, personalization of gaming experiences, and brain controlled-interfaces. In these cases, the utility to be preserved should be enough to provide a useful application, i.e., recognize the user, offer personalized options and responsive interfaces all with a tolerable error that does not hamper the security and usability of the service. 
 %Yao: EEG can be applied to diverse areas like disease identification [30], personal recognition [28], visual image generation using brainwaves [21], and brain typing [20]. 
 %TODO; \textcolor{magenta}{To be completed:  Neuromarketing? image of alcoholic vs non alcoholic brain waves to illustrate the case?}

%say something about the authentication accuracy so the potential for user identification is understood?
%
\subsubsection{Threat space} Brain activity is rich in information. It can be used to uniquely identify individuals given their unique characteristics and, in fact, several biometric systems based on brainwaves have been proposed~\cite{gui2019survey}. Besides, the acquisition of EEG signals raises privacy issues because brainwaves correlate, among others, with our mental states, cognitive abilities, and medical conditions~\cite{Sur2009}. \del{A third party in possession of neural data could try to make inferences of private attributes that were not intentionally disclosed by the user subscribing to its service, and thus non consented. Furthermore, if this entity has the ability to control the stimuli presented to the user when collecting their brainwave activity, such as the images shown in the computer screen, it could manipulate them to obtain private data.  Martinovic et al.~\cite{MDFP12}  where among the first to demonstrate the feasibility of these type of attacks.  Focusing on users of low cost EEG readers, they successfully proved that, by manipulating the images presented to the users, their EEG signals could reveal private information about, e.g., bank cards, PIN numbers, area of living, or if the user knew a particular person.}
\new{ Martinovic et al.~\cite{MDFP12} demonstrated that by manipulating the images presented to the users, their EEG signals could reveal private information, e.g., bank cards, PIN numbers, area of living, or if the user knew a particular person.} In another work, Frank et al.~\cite{Frank2013} show how to obtain private data from EEG recordings but, in this case, through subliminal stimuli (short duration images embedded in visual content) that cannot even be consciously detected by users.
\del{On the positive side, contrary to other behavioral traits like keystrokes or gait, brainwaves cannot be observed from the outside, which limits the possibility to misuse observed data to identify users without consent~\cite{korany2019xmodal}. Overall, the two main threats that apply to brainwave data collected/processed by a third party service provider are re-identification and inference of private attributes. In the first case, the attack would consist of linking the brain data of the user to brain data collected by other service or available in public databases, gaining additional information about the user that can potentially be identifiable or reveal sensitive information.  %profiling
     In the second case, the attack is oriented to uncover attributes correlated with the brainwave data, such as emotions, for which the user did not consent.}

\subsubsection{Anonymization Techniques}

\del{We found two works on brainwave data anonymization~\cite{matovu_jekyll_2018, yao2019improved}, both of them targeting the privacy goal of avoiding sensitive attribute %\tc{Which one?} 
inferences, more specifically, being  an alcoholic , through feature removal.}

\new{We found that all but one works for brainwave data anonymization rely on Generative Adversarial Networks (GANs) to protect the privacy by generating synthetic data. The large number of works published in the last two years indicates that anonymizing brain activity data is gaining some traction.}

\paragraph{Feature removal} Matovu et al.~\cite{matovu2016your} explore how to reduce the leakage of private information from EEG user authentication templates. They assume an insider type of attacker, such as an unscrupulous database administrator, who misuses their privilege to maliciously exploit the templates. The attacker wants to infer, specifically, if the user associated with a template is an alcoholic. Their envisioned anonymization technique aims at concealing the alcoholism information while still providing good authentication accuracy. It is, therefore, an attribute protection mechanism. Conceptually, it lies on the hypothesis that different template designs (features, channels, frequencies) will have an impact on the amount of non-authentication information (emotions, health conditions) that can be inferred. The authors demonstrate this hypothesis by choosing two different templates and calculating the predictive capability to authenticate users and to determine their alcohol consumption behavior. One of the template designs shows a good trade-off between accuracy and alcoholism obfuscation, while the other template provides better accuracy at the expense of leaking alcohol consumption behavior. While these results support the hypothesis, the article does not propose a concrete and systematic methodology to design the templates.

\paragraph{\new{Continuous Conversion}}
In the same direction of feature selection, Yao et al.~\cite{yao2019improved} propose the usage of Generative Adversarial Networks (GANs)~\cite{goodfellow2014generative} to filter sensitive information out of EEG data. Their goal is to reduce the possibility of inferring alcoholism while keeping the brain activity recordings useful to detect mental tasks, specifically to predict which visual stimulus the user is looking at. The GAN-based proposed filter involves deep neural networks that perform domain transformation, that is, translating EEGs from a source domain distribution X with both desired and privacy-related features to a target domain distribution Y with desired features only. Their results after applying the filtering technique show a significant reduction in the percentage of EEG sequences from alcoholic users that can be classified as such (from 90.6\% to 0.6\%). At the same time, the mental task classification accuracy does not drop significantly (4.2\% less). However, the original mental task classifier accuracy was not strong before filtering the privacy-sensitive features and it remains to be studied if this technique would work in other classification scenarios.

\new{
Pascual et al.~\cite{pascual_epilepsygan_2021} use a GAN to generate synthetic EEG data to train an epilepsy monitoring system as sharing large amounts of medical EEG is a privacy problem. The authors focus on inter-ictal EEG signals (signals between two seizures) as these are easier to record then the actual seizures. As generator a convolutional auto encoder is used but instead of decoding an inter-ictal the latent code is translated into an ictal sample. The descriminator then compares the synthetic ictal to a real. Their results show that the synthetic data reaches identification rate which are close to chance level, even when only two patients are in the test set. However, this is only a pseudonymization of the patients as all synthetic ictal values generated for a specific patient can still be linked to each other.\\
}
\new{
Bethge et al.~\cite{bethge_domain-invariant_2022} proposed privacy encoders to remove the sensitive information from each of the brain activity data streams before they are used in a classification task. For each data set a convolutional neural network is trained as encoder using the maximum mean discrepancy (MMD) between the different encoded data sets as loss function. This way the encoders should learn a domain-invariant representation of the data. They test their approach on four data sets finding that the classification from which data set a sample originated drops from 99\% to 52\%, while the emotion classification is only reduce from 51\% to 49\%. It remains an open question how well the identity of a subject would be preserved by this approach.
}

\new{
\paragraph{Continuous Conversion + Noise injection}
Debie et al.~\cite{debie_privacy-preserving_2020} also use a GAN to generate new synthetic data from the original one. They differ from Yao et al. and Pascual et al. in that they use differentially private stochastic gradient descent on the discriminator of the network. This method reduces the influence of each individual to the computation of the gradients. They evaluated their GAN on the Graz data set A with EEG data from 9 subjects. Their results show that the utility of the synthetic data is well preserved, however no additional privacy evaluation was performed.
}

\subsubsection{Evaluation}
The reviewed works, similar to the proposals for anonymizing gait, evaluate the quality of inference protection
by comparing the prediction accuracy for the protected attribute before and after modifying the EEG data.
The metrics used for this analysis are typical machine learning metrics, including accuracy, false positive rates, and false negative rates. Similarly, the loss of utility is evaluated by measuring the reduction in classification accuracy when using the original and anonymized EEG data. 

Both works used the same publicly available dataset for evaluating their anonymization proposals, the SUNY medical dataset with EEG data of 25 alcoholic subjects and 25 control subjects while looking at visual stimuli~\cite{EEGDatabase, karamzadeh2015relative}.

%\pc{to add: no other utility evaluated!! overhead (specially in the case of GANS and real time applicability)!, how to do research for other type of inferences?}

%\subsubsection{Limitations}

%+++++++++++++++++++++++
%EXTRA
%+++++++++++++++++++++
    
%MAybe for the discussion
%\textcolor{red}{
% With the potential wide adoption of BCI applications in our everyday lives, security and privacy concerns are rising~\cite{bonaci2014app, bernal2019cybersecurity}. Our user study and other previous research~\cite{futureAuthEEG} show that users are concerned about `mind reading’, but some people are already giving their brainwaves to third parties that offer brain-controlled games or relaxation applications. It is therefore paramount to research the security and privacy implications of using brainwaves in computer systems and work to design appropriate countermeasures before mainstream adoption.  }

%SCHEMA
%\textit{Utility}. 
%\textit{Threat Space.}
%\textit{Privacy Goals.}
%\textit{Anonymization Techniques.}
%\textit{Evaluation. Metrics and Datasets.}
%\textit{Limitations.} ??

%https://www.acm.org/publications/authors/submissions
%https://dl.acm.org/journal/csur/author-guidelines
%https://www.acm.org/publications/authors/reference-formatting

%% file: sections/table_waves.tex
\begin{table}[hbt!]
{\scriptsize
\addtolength{\tabcolsep}{-5pt}
\begin{tabular}{ |c|c|c|c| } 
 \hline
\begin{minipage}[c]{1.5cm} \centering \textbf{Wave Type} \end{minipage} & 
\begin{minipage}[c]{0.6cm} \centering \textbf{Freq. (Hz)} \end{minipage} & 
\begin{minipage}[c][0.7cm]{3cm} \centering \textbf{Originating Location} \end{minipage} & 
\begin{minipage}[c]{3.25cm} \centering \textbf{Mental State} \end{minipage} 
\\

 \hline\hline
 
\textbf{\textit{Gamma $\gamma$}} & 30-100 & Somatosensory cortex & \begin{minipage}[c][1cm]{3.25cm} \centering Active information processing, strong response to visual stimuli \cite{AAA15} \end{minipage}\\
\hline

\textbf{\textit{Beta $\beta$}} & 13-30 & \begin{minipage}[c][0.6cm]{3cm} \centering Both hemispheres, frontal lobe \end{minipage} & 
\begin{minipage}[c][0.7cm]{3.25cm} \centering Increased alertness, anxious thinking, focused attention \end{minipage}\\
\hline

\textbf{\textit{Alpha $\alpha$}} & 8-13 & \begin{minipage}[c][0.9cm]{3cm} \centering Posterior regions, both hemispheres; \\ High amplitude waves \end{minipage} & \begin{minipage}[c][1cm]{3.25cm} \centering Resting, eyes closed, no attention \cite{KSR12}; \\ Most dominant rhythm \end{minipage}\\
\hline

\textbf{\textit{Theta $\theta$}} & 4-8 & No special location & \begin{minipage}[c][0.7cm]{3.25cm} \centering Idling, dreaming, imagining, 
quiet focus, memory retrieval \end{minipage}\\
\hline 
 
\textbf{\textit{Delta $\delta$}} & 0.5-4 & \begin{minipage}[c][0.7cm]{2.3cm} \centering Frontal regions; \\High amplitude waves \end{minipage} & 
\begin{minipage}[c][0.7cm]{3.25cm} \centering Dreamless and deep sleep, unconsciousness \end{minipage}\\
\hline 
  
\end{tabular}
}
\\
\caption{Overview of EEG brainwaves - based on \cite{AE13} and \cite{AAA15}.}
\label{tab:waves}
\end{table}

%% file: sections/overview_table.tex
\begin{table*}[!ht]
\begin{center}
\begin{tabularx}{0.95\textwidth}{|p{3.0cm}|p{3.5cm}|X|X|X|X|X|}
\hline
\diagbox[innerwidth=3.0cm,height=1.0cm]{Method}{Trait} & Voice & Gait & Hand motion & Eye-Gaze & Heartbeat & Brain\newline activity\\
\hline
{\perturb} &  
\cite{parthasarathi_wordless_2013} \cite{mtibaa_cancelable_2018} &
\cite{hoang_gait_2015} &
\cite{maiorana_bioconvolving_2011} \cite{goubaru_consideration_2014} \cite{maiti_smartwatch-based_2016} \cite{vassallo_privacy-preserving_2017} &
\cite{david-john_for_2022} & \cite{HB7}$^\ast$& 
\\
\hline
{\noise} &
\cite{tamesue_sound_2014} \cite{hashimoto_privacy-preserving_2016} \cite{hamm_enhancing_2017} \cite{ohshio_active_2018} \cite{vaidya_you_2019} \cite{ma_you_2021} \cite{han_voice-indistinguishability_2020} &
\cite{tieu_approach_2017} \cite{tieu_spatio-temporal_2019} \cite{matovu_jekyll_2018} \cite{Understanding_Hanisch_2023} &
\cite{migdal_keystroke_2019} \cite{monaco_obfuscating_2017} \cite{shahid_evaluating_2021} & \cite{Steil19ETRA}\cite{bozkir2020differential} \cite{liu2019differential} \cite{Li_2021_Kaleido_USENIX} \cite{david-john_privacy-preserving_2021} \cite{hu_otus_2022} & &\\
\hline
{\coarse} & & & \cite{maiti_smartwatch-based_2016} \cite{vassallo_privacy-preserving_2017} & \cite{david-john_privacy-preserving_2021} & &\\
\hline
{\feature} & \cite{parthasarathi_speaker_2009} \cite{parthasarathi_lp_nodate} \cite{wyatt_conversation_nodate} \cite{zhang_privacy-preserving_2012} \cite{nelus_gender_2018}  \cite{cohen-hadria_voice_2019} \cite{ditthapron_privacy-preserving_2021} \cite{nelus_privacy-preserving_2021}  &
\cite{jourdan_toward_2018} \cite{garofalo_siamese_2020} \cite{Understanding_Hanisch_2023} & &  & \cite{HB12} \cite{HB13} \cite{HB17} & \cite{matovu2016your}  \cite{yao2019improved}\\
\hline
{\discrete} & 
\cite{pathak_privacy-preserving_2012} \cite{portelo_secure_nodate} \cite{portelo_privacy-preserving_2014} \cite{billeb_biometric_2015} &
 &
\cite{sae-bae_online_nodate}\cite{leinonen_preventing_2017}\cite{migdal_my_2019} \cite{vassallo_privacy-preserving_2017}\cite{figueiredo_prepose_2016} \cite{mukojima_deep-learning-assisted_2022} & &  \cite{HB32}$^\ast$&\\
\hline
{\cont} & 
\cite{jin_voice_2009}\cite{pobar_online_2014}\cite{justin_intelligibility_14}\cite{justin_speaker_2015}\cite{abou-zleikha_discriminative_2015}\cite{pribil_evaluation_2018}\cite{bahmaninezhad_convolutional_2018} \cite{fang_speaker_2019}\cite{keskin_measuring_2019}\cite{faundez-zanuy_speaker_2015}\cite{abad_advances_2016}\cite{lopezotero_influence_2017}\cite{magarinos_reversible_2017} \cite{aloufi_emotionless_2019}\cite{srivastava_evaluating_2020}\cite{gupta_design_nodate}\cite{aloufi_privacy-preserving_2020}\cite{yoo_speaker_2020}\cite{prajapati_voice_2021} \cite{ali_privacy_2021}\cite{patino_speaker_2021}\cite{mawalim_speaker_2022}\cite{canuto_effective_2014}$^\dagger$\cite{kondo_towards_2013}$^\ast$\cite{kondo_gender-dependent_2014}$^\ast$ \cite{qian_hidebehind_2018}$^\ast$\cite{qian_speech_2021}$^\ast$ \cite{srivastava_evaluating_2020}$^\ast$  &
\cite{agrawal_person_2011} \cite{ivasic-kos_person_2014} \cite{thapar_anonymizing_2021} \cite{hirose_anonymization_2019}$^\ddagger$ &
\cite{maiorana_bioconvolving_2011} \cite{malekzadeh_privacy_2020} \cite{saunders_anonysign_2021} \cite{xia_sign_nodate}&
\cite{david-john_for_2022} \cite{fuhl2021reinforcement} & 
\cite{bennis_application_2021}\cite{piacentino2020generating} \cite{HB5}$^\dagger$\cite{HB28}$^\dagger$ \cite{HB10}$^\dagger$\cite{HB24}$^\ast$ \cite{HB11}$^\ast$\cite{HB21}$^\ast$ \cite{HB8}$^\ast$ & \cite{pascual_epilepsygan_2021} \cite{bethge_domain-invariant_2022} \cite{debie_privacy-preserving_2020}$^\ast$\\
\hline
\end{tabularx}
\end{center}
\caption{An overview of all found methods classified by trait and method. Papers that propose multiple methods can appear in multiple rows. Papers that combine multiple methods are marked the following:  $^\ast$ plus noise injection, $^\dagger$ plus random perturbation,  $^\ddagger$ plus discrete conversion.}\label{table:techniques}
\end{table*}

\begin{table*}[!ht]
\begin{center}
\begin{tabularx}{0.95\textwidth}{|p{3.0cm}|p{3.5cm}|X|X|X|X|X|}
\hline
\diagbox[innerwidth=3.0cm,height=1.0cm]{Privacy Goal}{Trait} & Voice & Gait & Hand motion & Eye-Gaze & Heartbeat & Brain\newline activity  \\
\hline

Attribute &
\cite{mtibaa_cancelable_2018}\cite{hamm_enhancing_2017}\cite{ma_you_2021} \cite{pathak_privacy-preserving_2012}\cite{portelo_secure_nodate}\cite{portelo_privacy-preserving_2014} \cite{billeb_biometric_2015}\cite{pribil_evaluation_2018}\cite{faundez-zanuy_speaker_2015} \cite{aloufi_emotionless_2019}\cite{aloufi_privacy-preserving_2020}\cite{canuto_effective_2014} &
\cite{hoang_gait_2015} \cite{garofalo_siamese_2020} \cite{Understanding_Hanisch_2023} &
\cite{maiorana_bioconvolving_2011}\cite{goubaru_consideration_2014}\cite{maiti_smartwatch-based_2016} \cite{vassallo_privacy-preserving_2017}\cite{sae-bae_online_nodate}\cite{migdal_my_2019} \cite{maiorana_bioconvolving_2011} &
\cite{Steil19ETRA} \cite{bozkir_differential_2021} \cite{fuhl2021reinforcement} &
\cite{HB12}\cite{HB13} \cite{HB17}\cite{HB5} \cite{HB28}\cite{HB10}\cite{HB24} \cite{HB11}\cite{HB21} &
\cite{matovu2016your} \cite{yao2019improved} \cite{bethge_domain-invariant_2022} \cite{debie_privacy-preserving_2020} \\
\hline
Identity &
\cite{parthasarathi_wordless_2013}\cite{parthasarathi_lp_nodate}\cite{hashimoto_privacy-preserving_2016}\cite{hamm_enhancing_2017}\cite{ma_you_2021}\cite{ohshio_active_2018} \cite{vaidya_you_2019}\cite{han_voice-indistinguishability_2020}\cite{parthasarathi_speaker_2009}\cite{wyatt_conversation_nodate}\cite{zhang_privacy-preserving_2012}\cite{nelus_gender_2018} \cite{nelus_privacy-preserving_2021}\cite{cohen-hadria_voice_2019}\cite{jin_voice_2009}\cite{pobar_online_2014}\cite{justin_intelligibility_14}\cite{justin_speaker_2015} \cite{abou-zleikha_discriminative_2015}\cite{pribil_evaluation_2018}\cite{bahmaninezhad_convolutional_2018}\cite{fang_speaker_2019}\cite{mawalim_speaker_2022}\cite{prajapati_voice_2021} \cite{keskin_measuring_2019}\cite{abad_advances_2016}\cite{lopezotero_influence_2017}\cite{magarinos_reversible_2017}\cite{aloufi_privacy-preserving_2020}\cite{srivastava_evaluating_2020} \cite{ali_privacy_2021}\cite{yoo_speaker_2020}\cite{patino_speaker_2021}\cite{gupta_design_nodate}\cite{kondo_towards_2013}\cite{kondo_gender-dependent_2014} \cite{qian_hidebehind_2018}\cite{qian_speech_2021}&
\cite{matovu_jekyll_2018}\cite{tieu_approach_2017} \cite{tieu_spatio-temporal_2019}\cite{tieu_color_2020} \cite{jourdan_toward_2018}\cite{garofalo_siamese_2020} \cite{agrawal_person_2011}\cite{ivasic-kos_person_2014}\cite{thapar_anonymizing_2021} \cite{hirose_anonymization_2019} \cite{Understanding_Hanisch_2023}&
\cite{migdal_keystroke_2019}\cite{monaco_obfuscating_2017} \cite{shahid_evaluating_2021}\cite{leinonen_preventing_2017} \cite{figueiredo_prepose_2016}\cite{mukojima_deep-learning-assisted_2022} \cite{malekzadeh_privacy_2020}\cite{saunders_anonysign_2021} \cite{xia_sign_nodate}&
\cite{bennis_application_2021}\cite{piacentino2020generating} \cite{HB7} \cite{HB32} \cite{HB8} & 
\cite{david-john_for_2022}\cite{Steil19ETRA} \cite{bozkir_differential_2021}\cite{liu2019differential} \cite{david-john_privacy-preserving_2021}\cite{hu_otus_2022} \cite{Li_2021_Kaleido_USENIX}\cite{david-john_for_2022} \cite{fuhl2021reinforcement}&
\cite{pascual_epilepsygan_2021} \cite{debie_privacy-preserving_2020}\\

\hline
\end{tabularx}
\end{center}
\caption{An overview over which privacy goals the different techniques try to achieve.}\label{table:goals}
\end{table*}

%% file: sections/discussion.tex
All reviewed behavioral biometric traits have in common that they are captured as a time-series tracking the change of the trait over time. Most traits, such as gait, hand motions, voice, and eye gaze are overt traits that can be observed from a distance and do not require the participation of the subject. These traits are often captured as a byproduct for other recordings, for example, video recordings. EEG and ECG on the other hand are secret traits that can mostly only be recorded by directly attaching sensors to the subject to measure them. We found the most anonymization methods for voice and the least for EEG. For the traits touch, thermal, lip-facial, and motion we could not find any methods.

%Due to the missing requirement of user participation for the observable traits they are more prone to be abused for surveillance, or identity theft. Given that we, and others, upload/store a lot of info about ourselves, there is plenty of basis for making inferences. Therefore it is necessary to protect these traits more severely from being stolen or abused. 

The \textbf{utility} of these traits is very diverse and is mostly unique to each trait and the application using it. It ranges from utilities such as the naturalness of a motion to the intelligibility of utterances.

Regarding their \textbf{threat space}, the traits are similar to each other because the instances they are recorded are increasing with the pervasiveness of digital capturing devices such as smartphones and wearables in our everyday life. Wearables are of especial interest as they are attached to the subject and can therefore allow continuous capture of behavioral data. As our literature review has shown all traits can be used for both identity and attribute inference, which then can be used in a wide variety of privacy threats such as surveillance, identity theft, or private attribute inference. The privacy goals, identity protection, and attribute protection are also the same for all the traits. However, voice has an additional privacy goal in which the content of the speech should be made unintelligible.

For the \textbf{techniques} (see Table~\ref{table:techniques} and Table~\ref{table:goals}) that we reviewed, we found that most of them fall into the category of continuous conversion, followed by feature removal and noise injection. Next are random perturbation and discrete conversion, with most discrete conversion methods aiming at template protection. Coarsening is the category with the least amount of methods. 
We observe several differences for the categories of our taxonomy, for the removal methods we find that the removal is not directly reversible, however, due to the high redundancy in behavioral biometric data it still might be possible to reconstruct the removed data. For the conversion methods, we often observe that the parameter space for the anonymizations is often rather small, making it possible that an attacker can link clear and anonymized data by brute forcing the parameters when the anonymization technique is known. In general, we find that the reversibility of conversion techniques still has to be evaluated better. For noise injection techniques we find that the strong dependency both temporal and physiological is a problem since they can be used to filter out the noise.

With regard to the techniques providing differential privacy, we have observed that none of them can be used continuously over time without completely compromising user privacy. The reason lies in that the privacy budget is necessarily finite, which means, by the sequential composition property of differential privacy~\cite{McSherry09SIGMOD}, that it will be consumed completely at some time instant. Surprisingly, this appears to be in contradiction to the intended use of most of the applications where differential privacy is guaranteed, namely, continuous monitoring in healthcare scenarios, and identification and authentication services (which clearly are not single-use services). In that respect, the use of related privacy notions intended for continuous observations (e.g., $w$-event differential privacy~\cite{kellaris2014differentially}) may come in handy.

We made the observation that most methods do not manipulate the temporal aspect of their data. Notable exceptions are Hirose et al.~\cite{hirose_anonymization_2019} and Maiti et al.~\cite{maiti_smartwatch-based_2016}. Since all traits result in time series data manipulating the temporal order or time differences between events could lead to some general anonymization techniques which work for multiple traits.
For attribute protection we find anonymizing intrinsic attributes (e.g., age, sex) to be difficult as it is not clear which part of the behavioral data is relevant for these attributes.

Further, we noticed a lack of even basic understanding of users’ privacy awareness and concerns about behavioral privacy. These are necessary to design protection techniques that consider user needs and requirements.

\iflong 
\begin{table*}[!ht]
\begin{center}
\begin{tabularx}{0.95\textwidth}{|p{4cm}|X|X|X|X|}
\hline
Name & Participants & Published & Source & Trait \\
\hline
TIMIT & 630 & 1993  & \cite{TIMIT} & Voice\\
Albayzin & 164 & 1993 & \cite{albayzin} & Voice\\
YOHO & 137 & 1994 & \cite{YOHO} & Voice\\
BioSecureID & 400 & 2009 & \cite{BioSecureID} & Voice\\
Billeb et al. & 701 & 2014 & \cite{billeb_biometric_2015} & Voice \\
Librispeech & 1166 & 2015 & \cite{Librispeech}& Voice \\
RSR2015 & 300 & 2015  & \cite{RSR2015}& Voice\\
VoxCeleb & 1251 & 2018  & \cite{VoxCeleb} & Voice\\
CSTR VCTK Corpus & 110 & 2019 & \cite{vctk} &  Voice \\
VCC 2016 & 10 & 2016 & \cite{vcc2016} & Voice \\
\hline
CASIA-B  & 124 & 2005 & \cite{szhengICIP2011} & Gait\\
%%i3DPost & 8 & 2009  & \cite{gkalelis2009i3dpost} & Gait\\
BEHAVE & 125 & 2010 & \cite{blunsden2010behave} & Gait\\
OU-ISIR  & 200 & 2012 & \cite{Makihara_CVATN2012} & Gait\\
EPIC-Kitchens & 32 & 2020 & \cite{Damen2020Collection} & Gait\\
IITMD-WFP & 31 & 2021 & \cite{thapar_anonymizing_2021} & Gait\\
\hline
MCYT baseline corpus  & 330 & 2003 & \cite{ortega-garcia_mcyt_2003} & Hand motion\\
SVC2004 & 100 & 2004 & \cite{kanade_svc2004_2004} & Hand motion \\
GREYC  & 133 & 2009 & \cite{giot_greyc_2009} & Hand motion\\
MNIST  & 500 & 2012 & \cite{li_deng_mnist_2012} & Hand motion\\
Web-based keystroke  & 83 & 2012 & \cite{giot_web-based_2012} & Hand motion\\
SMILE & 30 & 2018 & \cite{SMILE} & Hand motion\\
ASLLRP & 33 & 2022 & \cite{ALSVideo}  & Hand motion\\ 
\hline
DGaze & 22 & 2020 & \cite{isha2020iros} & Eye-Gaze\\
%%ET-DK2 & 18 & 2021 & \cite{brendan_david_john_2021_4642612} & Eye-Gaze\\
VR-Saliency & 169 & 2018 & \cite{8269807} & Eye-Gaze\\
Gaze Prediction & 43 & 2018 & \cite{8578657} & Eye-Gaze\\
Video viewing & 50 & 2017 & \cite{VideoViewing} & Eye-Gaze\\
MPIIDPEye & 20 & 2019 & \cite{steil19_etra_2} & Eye-Gaze\\
OpenEDS & 157 & 2019 & \cite{OpenEDS} & Eye-Gaze\\
EHTask & 30 & 2022 & \cite{hu22_EHTask} & Eye-Gaze\\
DOVES & 29 & 2009 & \cite{doves} & Eye-Gaze\\
\hline
SUNY EEG database  & 50 & 1999 & \cite{EEGDatabase} & Brain activity\\
UCI EEG database & 122 & 1999 & \cite{UCIEEG} & Brain activity\\
DEAP & 32 & 2011 & \cite{5871728} & Brain activity\\
DREAMER & 23 & 2018 & \cite{katsigiannis_stamos_2017_546113} & Brain activity\\
SEED & 15 & 2015 & \cite{zheng2015investigating} & Brain activity\\
\hline
MIT-BIH ECG Arrhythmia & 47 & 1979 & \cite{MIT-DB} & Heartbeat\\
%%MIT-BIH Noise Stress Test & 2 & 1984 & \cite{MIT-SN} & Heartbeat\\
Physikalisch Technische Bundesanstalt & 290 & 1995 & \cite{PTB-DB} & Heartbeat\\
\hline
\end{tabularx}
\end{center}
\caption{An overview of used behavioral biometric datasets.}\label{table:datasets}
\end{table*} 
\else \fi

We found that the evaluation methodology between the traits and methods is rather similar. In general, an inference/recognition system is being used on the clear and on the anonymized data and then the difference in accuracy is reported, often without retraining the inference system on the anonymized data. We find this methodology too simple as the underlying assumption is that the attacker is not aware of the anonymization. Besides training the recognition model on the anonymized data the evaluations should also consider an attacker that actively tries to reverse the anonymization and knows the anonymization technique and its parameters. To allow the comparison between multiple methods the attacker models should be made explicit and common, similar to attacker models in cryptography. 
Only a handful of papers compare their own methods to that of others and due to the differences in attacker models and data sources, it is difficult to compare their results to one another. We also found that there are not many approaches~\cite{zhang_enhancing_2020, qian_towards_2018} to formalize the privacy of behavioral biometric anonymization methods and most of the evaluations rely on empirical privacy estimations. Another problem is that the evaluation methodology is too close to the recognition system evaluation methodology which seeks to infer persons in a large dataset with poor data quality, while an anonymization method should also work on a small group size with high data quality. We believe that the lack of available datasets \iflong(see Table~\ref{table:datasets})\else\fi is one of the main problems which keeps the less researched behavioral biometric traits back.